\documentclass[11pt]{article}
\usepackage{amssymb}
\usepackage{amsmath}
\usepackage{amstext}
\usepackage{graphicx,epsfig}
\usepackage{epsfig}
\usepackage{verbatim} 
\usepackage{fancybox}
\usepackage{color}
\usepackage{ulem}
\usepackage{enumitem}
\usepackage{subfigure}
\usepackage{bbm}
\usepackage{parskip}
\usepackage{dsfont}
\usepackage[numbers,sort&compress]{natbib}

\newcommand{\Comment}[1]{{}}
\definecolor{darkblue}{rgb}{0.15,0.35,0.55}
\definecolor{reddish}{rgb}{0.65, 0.2, 0.2}
\usepackage[linktocpage=true]{hyperref}
\hypersetup{
colorlinks=true,
citecolor=darkblue,
linkcolor=reddish,
urlcolor=darkblue,
pdfauthor={},
pdftitle={},
pdfsubject={}
}

\newcommand{\be}{\begin{equation}}
\newcommand{\ee}{\end{equation}}
\newcommand{\bea}{\begin{eqnarray}}
\newcommand{\eea}{\end{eqnarray}}
\newcommand{\beas}{\begin{eqnarray*}}
\newcommand{\eeas}{\end{eqnarray*}}
\newcommand{\nn}{\nonumber \\ }

\def\({\left(}
\def\){\right)}

\newcommand{\rd}{{\rm d}}

\def\gsim{ \lower .75ex \hbox{$\sim$} \llap{\raise .27ex \hbox{$>$}} }
\def\lsim{ \lower .75ex \hbox{$\sim$} \llap{\raise .27ex \hbox{$<$}} }

\title{}
\author{}

\numberwithin{equation}{section}

\begin{document}
%
~
\vspace{1truecm}
\renewcommand{\thefootnote}{\fnsymbol{footnote}}
\begin{center}
{\huge \bf{Einstein Gravity, Massive Gravity, \\  \vspace{7pt} Multi-Gravity and Nonlinear Realizations}}
\end{center} 

\vspace{1truecm}
\thispagestyle{empty}
\centerline{\Large Garrett Goon,${}^{\rm a,}$\footnote{\tt gg399@cam.ac.uk} Kurt Hinterbichler,${}^{\rm b,}$\footnote{\tt khinterbichler@perimeterinstitute.ca} Austin Joyce${}^{\rm c,}$\footnote{\tt ajoy@uchicago.edu} and Mark Trodden${}^{\rm d,}$\footnote{\tt trodden@physics.upenn.edu}}
\vspace{.7cm}

\centerline{\it ${}^{\rm a}$Department of Applied Mathematics and Theoretical Physics}
\centerline{\it Cambridge University, Cambridge, CB3 0WA, UK}

\vspace{.3cm}

\centerline{\it ${}^{\rm b}$Perimeter Institute for Theoretical Physics}
\centerline{\it 31 Caroline St. N., Waterloo, ON, N2L 2Y5, Canada}

\vspace{.3cm}

\centerline{\it ${}^{\rm c}$Enrico Fermi Institute and Kavli Institute for Cosmological Physics}
\centerline{\it University of Chicago, Chicago, IL 60637, USA}
\vspace{.3cm}
\centerline{\it$^{\rm d}$Center for Particle Cosmology, Department of Physics and Astronomy,}
\centerline{\it University of Pennsylvania, Philadelphia, PA 19104, USA}

\vspace{.5cm}
\begin{abstract}
\vspace{.03cm}
\noindent
The existence of a ghost free theory of massive gravity begs for an interpretation as a Higgs phase of General Relativity. We revisit the study of massive gravity as a Higgs phase. Absent a compelling microphysical model of spontaneous symmetry breaking in gravity, we approach this problem from the viewpoint of nonlinear realizations.  We employ the coset construction to search for the most restrictive symmetry breaking pattern whose low energy theory will both admit the de Rham--Gabadadze--Tolley (dRGT) potentials and nonlinearly realize every symmetry of General Relativity, thereby providing a new perspective from which to build theories of massive gravity. In addition to the known ghost-free terms, we find a novel parity violating interaction which preserves the constraint structure of the theory, but which vanishes on the normal branch of the theory.  Finally, the procedure is extended to the cases of bi-gravity and multi-vielbein theories.  Analogous parity violating interactions exist here, too, and may be non-trivial for certain classes of multi-metric theories.
\end{abstract}

\newpage

\tableofcontents

\newpage
\renewcommand*{\thefootnote}{\arabic{footnote}}
\setcounter{footnote}{0}

\section{Introduction} 

The long standing problem of constructing a ghost-free interacting theory of a massive spin-2 field has been solved in the last few years by de Rham, Gabadadze and Tolley (dRGT)~\cite{deRham:2010ik,deRham:2010kj}.  It is strongly tempting to interpret this theory as a Higgs phase of General Relativity (GR), but to date a satisfactory microscopic theory which leads to Poincar\'e-invariant massive gravity via symmetry breaking remains elusive.\footnote{The ghost condensate \cite{ArkaniHamed:2003uy} can be thought of as a Higgs-ing that leads to Lorentz violation. Another recent proposal for a UV extension of Lorentz-violating massive gravity is~\cite{Blas:2014ira}.  For an intriguing construction on AdS space, see~\cite{Porrati:2001db,Porrati:2003sa}.} Therefore, to investigate gravity in a Higgs-ed phase, one is forced to take a somewhat broader view and focus on gross features which follow from symmetry breaking. The goal of this paper is to revisit the approach of viewing massive gravity as a theory of spontaneously broken gauge symmetry by employing nonlinear realization techniques to construct the effective theory. 

Before undertaking the construction of massive gravity in this language, we first consider constructing Einstein gravity itself as a theory of a Goldstone field, the metric, which nonlinearly realizes an infinite number of rigid symmetries making up the diffeomorphism symmetry. This viewpoint has been considered before~\cite{Ogievetsky:1973ik,Borisov:1974bn,Ivanov:1981wn,Pashnev:1997xk,Riccioni:2009hi,Delacretaz:2014oxa}, but our approach to General Relativity differs in that it constructs the theory in the vielbein formulation and explicitly nonlinearly realizes both diffeomorphisms and local Lorentz transformations---this may prove to be useful for some applications, and is crucial for the later construction of massive gravity. Similar techniques can be employed to construct Yang--Mills gauge fields as Goldstone bosons~\cite{Ivanov:1976pg}. In~\cite{Goon:2014ika} it was shown that these techniques can be extended to treat gauge theories in a spontaneously broken phase, leading to a theory of massive spin-1 fields (or equivalently gauge fields and St\"uckelberg fields). It is this approach we will employ here: after constructing Einstein gravity from the symmetry-breaking viewpoint, we will the proceed to consider ``Higgs-ing" the construction in order to build the effective theory for the broken phase. Completely analogously to the Yang--Mills case, we find that the construction leads to the theory of a massive spin-2 field.

Our aim is more than the construction of an arbitrary theory of a massive spin-2 particle, since it is known that a generic such theory will propagate an additional, ghostly, polarization~\cite{Boulware:1973my}, as we  review below. Thus, we are motivated to see whether there is a symmetry breaking pattern for which the low energy degree of freedom is a massive spin-2 and for which {\it only} the dRGT potentials are allowed.  As we will argue, this appears not to be possible.  This is perhaps not surprising, as it appears that quantum corrections in the dRGT theory generate additional terms~\cite{deRham:2013qqa}.  Another possibility is that the dRGT terms could be singled out as Wess--Zumino terms of the construction, as has been shown to happen for the Galileon scalar field theories that emerge as a particular limit of dRGT~\cite{Goon:2012dy}.  This does not happen either.   However, there are a number of intriguing features that appear in the course of the construction, which we summarize here:

\begin{itemize}

\item The dRGT potentials appear as among the simplest terms that can be constructed, requiring only the exterior product of forms. Similarly, in the Einstein gravity case, the Lovelock invariants are the only terms that may be constructed in this way. This reinforces the notion that the dRGT potentials are in some sense generalized characteristic classes.

\item In addition to the dRGT terms, we identify a parity-violating term that appears to be unstudied in the literature. This term depends only on the anti-symmetric part of the vielbein, and so it vanishes on the usual branches of dRGT and bi-metric gravity, where symmetry of the vielbein is important to the equivalence of the metric and vielbein formulations. However, in non-trivial branches of in multi-metric situations where the theory graphs of \cite{Hinterbichler:2012cn} form closed loops, this term may possibly be nontrivial. 

\end{itemize}

In the remainder of this section, we briefly review the construction of ghost-free massive gravity and comment on the idea of Higgs-ing gravity and its relation to massive gravity in general. We then review the viewpoint of gauge fields as Goldstone bosons, which is somewhat non-standard but is a powerful formal tool. In Section~\ref{cosetconst}, we review the coset construction formalism, in which the remainder of the paper is cast. In Section~\ref{grconstruction} we apply this formalism to Einstein gravity, after identifying the appropriate infinite-dimensional symmetry algebras to be nonlinearly realized. Next, we consider further breaking of the symmetries in Section~\ref{massivegconstruct} in order to construct massive gravity.  We then apply the same techniques in Section~\ref{biandmultig} to the closely related bi-gravity and multi-vielbein theories, which follow straightforwardly. Finally in Section~\ref{conclusion} we conclude and provide some remarks about applications and insights from this construction.

 \subsection{Brief review of massive gravity}
The history of massive gravity is long and complicated and there are still unresolved issues in the field (see e.g. \cite{Deser:2014fta} and the counterarguments in \cite{deRham:2014zqa}), so we only review the bare minimum required here (see the reviews~\cite{Hinterbichler:2011tt,deRham:2014zqa} for more details).

Fierz and Pauli first wrote down a consistent quadratic theory which propagates the appropriate five degrees of freedom of a massive spin-2 field of mass $m$~\cite{Fierz:1939ix},
 \bea
  \mathcal{L}= -\frac{1}{2}\partial_\lambda h_{\mu\nu}\partial^\lambda h^{\mu\nu}+\partial_\mu h_{\nu\lambda}\partial^\nu h^{\mu\lambda}-\partial_\mu h^{\mu\nu}\partial_\nu h+\frac{1}{2}\partial_\lambda h\partial^\lambda h 
-\frac{1}{2}m^{2}\left (h_{\mu\nu}h^{\mu\nu}-h^{2}\right )\ ,\label{FierzPauliTheory}
\eea
 where $h_{\mu\nu}$ is the metric perturbation about flat space, $g_{\mu\nu}=\eta_{\mu\nu}+h_{\mu\nu}$.  The two-derivatives terms in \eqref{FierzPauliTheory} comprise the standard quadratic kinetic term---which is the Einstein--Hilbert term of GR linearized about flat space (all indices are raised and lowered with $\eta_{\mu\nu}$, and $h\equiv h_{\mu\nu}\eta^{\mu\nu}$). Note that the mass terms $\sim m^2h^2$ break diffeomorphism invariance.  While~\eqref{FierzPauliTheory} provides a consistent starting point, Boulware and Deser showed that the addition of interaction terms generically causes the theory to propagate a $6^{\rm th}$ degree of freedom, which is a ghost~\cite{Boulware:1973my}.
  
However, de Rham, Gabadadze and Tolley (dRGT), building from the results of \cite{ArkaniHamed:2002sp,Creminelli:2005qk}, discovered that, by suitably tuning the interactions, one can form a specific nonlinear theory which continues to propagate only five degrees of freedom~\cite{deRham:2010kj,deRham:2010ik}.  The absence of the sixth degree of freedom was shown conclusively in~\cite{Hassan:2011hr,Hassan:2011ea} through a Hamiltonian analysis. The original formulation of the theory involves intricate potentials built from the square root matrix $(\sqrt{g^{-1}\eta})^{\mu}{}_{\nu}$. However, for our purposes an alternative formulation of the dRGT theory in terms of vielbeins is more useful and is given in\footnote{The vielbein formulation of dRGT massive gravity was anticipated in~\cite{Zumino:1970tu,Nibbelink:2006sz}.} \cite{Hinterbichler:2012cn}, whose conventions we follow. Given the vielbeins $e_{\mu}{}^{a}$, related to the metric via $g_{\mu\nu}=\eta_{ab}e_{\mu}{}^{a}e_{\nu}{}^{b}$, and vielbein one-forms ${\bf e}^{a}\equiv e_{\mu}{}^{a}\rd x^{\mu}$, the dRGT action can be written in the following form 
\be
S = \frac{M_{\rm pl}^{2}}{2}\int\rd^{4}x\det (e)\,R[e]-\frac{m^{2}M_{\rm pl}^{2}}{8}\int\mathcal{L}_{m}^{\rm dRGT}~,\label{dRGTAction}
\ee
where the dRGT mass terms are given by
\begin{align}
\nonumber
\mathcal{L}_{m}^{\rm dRGT}= ~~\frac{ \beta_{0}}{4!}\epsilon_{abcd}{\bf e}^{a}\wedge{\bf e}^{b}\wedge{\bf e}^{c}\wedge{\bf e}^{d}
&+\frac{\beta_{1}}{3!}\epsilon_{abcd}{\bf 1}^{a}\wedge{\bf e}^{b}\wedge{\bf e}^{c}\wedge{\bf e}^{d}\\
+\frac{\beta_{2}}{4}\epsilon_{abcd}{\bf 1}^{a}\wedge{\bf 1}^{b}\wedge{\bf e}^{c}\wedge{\bf e}^{d}
&+\frac{\beta_{3}}{3!}\epsilon_{abcd}{\bf 1}^{a}\wedge{\bf 1}^{b}\wedge{\bf 1}^{c}\wedge{\bf e}^{d}\ .\label{dRGTMassTerms}
\end{align}
In \eqref{dRGTMassTerms}, ${\bf 1}^{a}\equiv\delta^{a}_{\mu}\rd x^{\mu}$ is the unit vielbein corresponding to the fixed Minkowski fiducial metric, and the $\beta_{0}$ term simply corresponds to the cosmological constant. In order for flat space to be a solution, the coefficients must satisfy $\beta_{0}+3\beta_{1}+3\beta_{2}+\beta_{3}=0$ and for the action~\eqref{dRGTAction} to correspond to a graviton of mass $m$ requires $\beta_{1}+2\beta_{2}+\beta_{3}=8$.   These specially-chosen potentials lead to the additional constraints that are necessary to exorcise the ghostly sixth degree of freedom that afflicts generic massive gravity theories. A primary goal of this paper is to determine under which conditions spontaneous symmetry breaking can lead to the action \eqref{dRGTAction}.
 
\subsection{Higgs-ing gravity} 

Since the discovery of spontaneous symmetry breaking (SSB), there have been numerous attempts to marry this phenomenon to gravity, for example by interpreting GR itself as the result of SSB \cite{Ivanenko:1984vf,Kirsch:2005st,Boulanger:2006tg} or by using a Higgs mechanism to give the graviton a mass \cite{'tHooft:2007bf,Chamseddine:2010ub} (see also \cite{Percacci:1984ai,Percacci:1990wy,Moffat:1992bf,Percacci:2009ij}).  

Focusing on the latter possibility, it is well-known that  if the graviton acquires a mass through SSB, the mechanism must be {qualitatively} different from the manner in which Yang--Mills gauge bosons become massive \cite{Kostelecky:1989jw}.  Schematically, a typical matter field, $\psi$, gauged under some group, $G$, will couple to the associated gauge bosons, $A_{\mu}$, through the covariant derivative so that the Lagrangian will contain a term $\mathcal{L}\supset \left ((\partial+A)\psi\right )^{2}$.  If $\psi$ then acquires a vacuum expectation value (VEV) via symmetry breaking, {\it i.e.}, $\langle\psi\rangle\neq 0$, this generates a mass term for the gauge bosons: $\mathcal{L} \supset \langle\psi\rangle^{2}A^{2}$.  
The directly analogous scenario for gravity does not generate a mass for the graviton because the gravitational covariant derivative causes $\psi$ to couple \textit{derivatively} to the graviton.  For instance, if $\psi$ has a Lorentz index then $\psi$ couples to the spin connection $\omega_{\mu}{}^{ab}$ through $\nabla\psi=(\partial+\omega)\psi$; since the spin connection involves derivatives of the gravitational field, a VEV for $\psi$ does not cause the gravitational covariant derivative to generate mass terms. The gravitational field {\it does} couple non-derivatively to the potential for $\psi$ through $\mathcal{L}\supset\sqrt{-g}\, V(\psi)$, but here, of course, a VEV for $\psi$ only leads to a cosmological constant, not a true mass term.  

Therefore, while a Higgs mechanism for gravity may exist, we expect that its form will be quite distinct from examples familiar to us from the study of Yang--Mills (some arguments suggest that if a Lorentz invariant UV completion does exist, it must be strongly coupled or be somehow non-field theoretic \cite{Adams:2006sv}).  Lacking a satisfactory microscopic model to examine, it is fruitful to turn to the more general features of symmetry breaking and focus on the generic properties that a potential model must display.  In particular, regardless of the details of the breaking mechanism, after a symmetry is spontaneously broken, the theory remains invariant under the symmetry, albeit in a nonlinearly-realized form. This places strict constraints on the resulting interactions of the low energy theory.
 
 \subsection{The coset construction and gauge theories}
 
To a remarkable extent, the physics of a system can be deduced from knowing the pattern of symmetry breaking. For example, consider a symmetry group, $G$, broken down to one of its subgroups, $H$. The broken phase will generically linearly realize the preserved symmetry subgroup $H$ and nonlinearly realize the elements $G/H$ that were not preserved in the breaking. Having specified such a pattern, Callan, Coleman, Wess and Zumino (CCWZ)~\cite{Coleman:1969sm,Callan:1969sn}, and independently Volkov~\cite{volkov}, developed a method through which one can algorithmically construct the most general Lagrangians which linearly realize $H$ and nonlinearly realize all the broken transformations. This so-called ``coset construction" captures the dynamics of the Goldstone modes which govern the low energy physics after spontaneous symmetry breaking.  

Such methods prove invaluable in the construction of an effective field theory (EFT) which describes the low energy physics.  These techniques are particularly powerful when the dynamics that lead to symmetry breaking are difficult to understand analytically, or are unknown. For example, in the case of pion physics~\cite{Weinberg:1968de}, strong dynamics break the approximate chiral symmetry of QCD---which is difficult to treat analytically---but the coset construction gives us access to an EFT in which low-energy quantities can be systematically calculated.  Even without a full understanding of the underlying theory, nonlinear realization techniques can provide us with non-trivial information regarding the broken phase.  We aim to apply these methods to understand gross features of a Higgs phase of gravity, despite our ignorance of the microscopic dynamics which generate the SSB.

Although the coset construction was initially developed to treat spontaneously broken {\it global} symmetries, this is not the only situation in which it is applicable. In particular, gauge theories nonlinearly realize the local versions of their symmetry groups and the coset construction can be employed here as well.  More precisely, the schematic transformation for a Yang--Mills gauge field is $A\mapsto U^{-1}(x)(A+\rd)U(x)$ which represents a linear transformation when $U(x)$ is global ({\it i.e.}, independent of $x$) and a nonlinear one when it is local.   Applying the coset methods to this scenario in which we take a typical YM gauge group $G$---say $SO(N)$ or $SU(N)$---and demand that global and local $G$ transformations are linearly and nonlinearly realized, respectively, one finds that the resulting building blocks are those of Yang--Mills theory~\cite{Ivanov:1976pg,Borisov:1974bn}. That is, one finds that the Lagrangian for the gauge field must be constructed from the usual field strength tensor $F_{\mu\nu}^{a}$ and that it couples to matter through the gauge covariant derivative, $\nabla=\partial+A$.  Therefore, starting only with a specification of the symmetries of the system, the coset construction picks out Yang--Mills as the proper representation.\footnote{Because gauge theories involve spacetime dependent transformations there are additional subtleties which do not arise in the internal symmetry case, see \cite{Goon:2014ika} for details.}  
 
In \cite{Goon:2014ika} it was demonstrated that coset methods also faithfully reproduce the physics of gauge theories in the Higgs phase.  In this scenario \textit{both} the global and local parts of the Yang--Mills gauge group $G$ are nonlinearly realized and one imagines that the nonlinear realization of global transformations is caused by some physical symmetry-breaking process. Applied here, coset methods determine that the low energy physics is governed by massive gauge bosons with mass terms written in the St\"uckelberg language, as expected.
 
 The central goal of this paper is to perform the analogous procedure for the case of gravity, beginning by constructing GR through the method of nonlinear realizations by identifying the appropriate groups and cosets required.\\
 \noindent
 {\bf Conventions:} We work with the mostly-plus metric signature $\eta_{\mu\nu} = (-,+,+,+,\cdots)$ throughout. We (anti)-symmetrize tensors with weight one {\it i.e.}, $S_{(\mu\nu)} = \frac{1}{2}(S_{\mu\nu}+S_{\nu\mu})$. The Levi--Civita symbol is defined so that $\epsilon_{0123\cdots}= +1$.

 \section{Review of nonlinear realizations}
 \label{cosetconst}
 
 Before building GR, we quickly review the coset construction algorithm.  In addition to the CCWZ formalism, we need to take into account some subtleties which arise in the case of spacetime symmetry breaking~\cite{Volkov:1973vd,Ogievetsky:1974,Ivanov:1975zq}.  For a more extensive discussion of the methods of nonlinear realizations, we refer the reader to Section 2 of~\cite{Goon:2012dy}.
 
 The CCWZ formalism begins by specifying a breaking pattern from some Lie group of symmetries, $G$, down to one of its subgroups, $H$,
 \be G\longrightarrow H.
 \ee
Let $\{V_{I}\}$ represent the generators of $H$, $\{Z_{a}\}$ represent the remaining---broken---generators and assume that the commutator of an element of $\{V_{I}\}$ with an element of $\{Z_{a}\}$ will never contain another $V_{I}$ type generator.\footnote{For a compact Lie algebra, it can be proven that bases can be chosen such that this is true, but for more general algebras, we take it as a simplifying assumption.} A canonical representative element of the coset $G/H$ is then written as $g(\xi)\equiv \exp \left(\xi^{a}Z_{a}\right)$ where the $\xi^{a}$'s correspond to Goldstone fields.  An arbitrary element $g'\in G$ generates a unique transformation $g(\xi)\mapsto\tilde{g}(\xi,g')$, with $\tilde{g}(\xi,g')$ defined via the condition
 \begin{align}
 g'g(\xi)=\tilde{g}(\xi,g')h(\xi,g')~,~~~~~~~~{\rm where}~~~ h(\xi,g)\in H\ .\label{MCtransformationcondition}
 \end{align}  Defining the fields $\tilde\xi^{a}$ by $\tilde{g}(\xi,g')\equiv \exp(\tilde\xi^{a}Z_{a} )$, the relation between $\tilde\xi^{a}$ and $\xi^{a}$ will be linear if $g'$ is an element of $H$, but is complicated and nonlinear otherwise.  In this manner the Goldstone fields, $\xi^{a}$, linearly realize the symmetries associated with the preserved subgroup $H$ and nonlinearly realize the remaining broken symmetries.
 
 In order to construct actions we first build the Maurer--Cartan (MC) form $\Omega\equiv g^{-1}(\xi)\rd g(\xi)$, which is a Lie algebra-valued 1-form.  Decomposing $\Omega$ into its parts along the broken and unbroken generators as
 \begin{align}
 \Omega&\equiv \Omega^{a}Z_{a}+\Omega^{I}V_{I}\equiv \Omega_{Z}+\Omega_{V}\ ,
 \end{align}
we find that under the transformation induced by an arbitrary element $g'\in G$ the components of the MC form transform as \begin{align}
 g':\begin{cases} \Omega_{Z}&\longmapsto ~h(\xi,g')\Omega_{Z}h^{-1}(\xi,g'),\\
\Omega_{V}&\longmapsto ~ h(\xi,g')\left (\Omega_{V}+\rd\right )h^{-1}(\xi,g').\end{cases}\ \label{MCTransformation}
 \end{align}
 The utility of the MC form is precisely that it has these nice transformation properties under the action of the group $G$. We can build Lagrangians which are invariant under \textit{all} the symmetries of $G$ by combining factors of $\Omega_{Z}$ together and tracing over group indices such that the resulting operator is invariant under the transformation~\eqref{MCTransformation}, {\it i.e.}, are $H$-invariant.  The $\Omega_{V}$ components transform as a connection and can be used to couple the Goldstone fields to other matter fields which transform in some representation of $H$.
 
An important subtlety that arises in the case of nonlinearly realized spacetime symmetries is the removal of fields via inverse Higgs (IH) constraints~\cite{Ivanov:1975zq}. For a symmetry breaking pattern $G\to H$, there are na\"ively $\dim(G/H)$ fields $\{\xi^{a}\}$ in the representative element $g$, which is the appropriate number of Goldstone modes for the case of internal symmetry breaking. However, it is well known that there can be fewer than $\dim(G/H)$ independent Goldstone degrees of freedom when spacetime symmetries are involved~\cite{Low:2001bw}.  In practice, the rule is that if the commutator between a preserved translation generator, $P_{\mu}$, and broken generator, $Z_{1}$, contains a second broken generator, $Z_{2}$---schematically $[P_{\mu},Z_{1}]\supset Z_{2}$---then is is possible to eliminate the Goldstone field corresponding to $Z_{1}$ by setting some part of the MC component along $Z_{2}$ to zero. Our perspective will be that the inverse Higgs constraints provide a mechanism through which one can consistently reduce the number of fields in the theory while still realizing all of the symmetries contained in $G$; whether or not to impose them is a choice.\footnote{The circumstances under which one is \textit{required} to impose an inverse Higgs constraint and eliminate the $Z_{1}$ field is still a matter of current research, see e.g.~\cite{McArthur:2010zm,Nicolis:2013sga,Brauner:2014aha}.}

Finally, if there are preserved translation generators, $P_{\mu}$, when treating a case of spacetime symmetry breaking, these generators are nevertheless treated on the same footing as the broken generators due to the fact that translations are nonlinearly realized on the spacetime coordinates. In this case the MC form is written
\be
\Omega =\Omega^{a}Z_{a}+\Omega^{I}V_{I}+\Omega^{\mu}P_{\mu}~.
\ee
In addition to constructing invariant actions by using the wedge product to combine forms, it is also possible to form a covariant derivative from these objects. 
In this case, the components along $P_{\mu}$ define a vielbein, $e_{\nu}{}^{\mu}$, via $\Omega^{\mu}P_{\mu}\equiv \rd x^{\nu}e_{\nu}{}^{\mu}P_{\mu}$.  This vielbein is in turn used to define an invariant measure, and a covariant derivative of the Goldstone fields through $\Omega^{a}Z_{a}=\rd x^{\nu}e_{\nu}{}^{\mu}\mathcal{D}_{\mu}\xi^{a}Z_{a}$.  Invariant actions are formed by contracting factors of covariant derivatives in an $H$-invariant fashion, and integrating with the invariant measure.

 \section{General relativity}
 \label{grconstruction}
 
We now turn to the construction of Einstein gravity using coset methods. In this construction, the graviton itself plays the role of a Goldstone field which nonlinearly realizes an infinite number of symmetries. This is possible because we can think of a gauge symmetry as an infinite number of rigid global symmetries, most of which are realized nonlinearly. This viewpoint has been explored multiple times before through a variety of slightly differing methods~\cite{Ogievetsky:1973ik,Borisov:1974bn,Pashnev:1997xk,Riccioni:2009hi,Ivanov:1981wn}. In particular, some aspects of this section are similar to Sec. 3 of the recent paper~\cite{Delacretaz:2014oxa}.  However, an important distinction between our construction and others' is that we include {both} the diffeomorphisms and local Lorentz transformation groups in the coset, which is important for the later construction of massive gravity. There are close parallels between the coset construction of GR and that of Yang--Mills in~\cite{Goon:2014ika}, which the reader may find helpful as background for the following sections.
 
 \subsection{Symmetries and algebras}
We first identify the algebra of symmetries which are realized, both linearly and nonlinearly, in Einstein gravity. We work in arbitrary $(d+1)$-dimensions; both Greek and Latin indices run over $\{0,1,\ldots,d\}$. In order to make contact most easily with massive gravity later, it will prove useful to work in the vielbein formalism. In this case, Einstein gravity is invariant under both spacetime diffeomorphisms, and local Lorentz transformations (LLT) which act on the tangent space (the vielbein indices).

The group we will consider is a slight extension of the group of diffeomorphisms plus LLTs; it is a semi-direct product of diffeomorphisms and an internal, local copy of the Poincar\'e group.  The extra internal translation generators will act trivially on the familiar vielbein and spin connection, but will be needed to obtain the correct fields in the coset construction.  Analogously to the treatment of gauge symmetries in~\cite{Goon:2014ika}, we expand the diffeomorphisms in powers of the spacetime coordinate and treat them as an infinite number of global symmetries generated by the set $P^{\nu_1\ldots\nu_{n}}{}_{\mu}$, where $n\in\{0,1,\ldots\}$, which are modeled by $P^{\nu_1\ldots\nu_{n}}{}_{\mu}=-x^{\nu_{1}}\ldots x^{\nu_{n}}\partial_{\mu}$ and which satisfy the commutation relations\footnote{Here and throughout, vertical bars around indices indicate that the enclosed indices are omitted from the (anti-)symmetrization.}
\be
  \left [P^{\mu_1\ldots\mu_m}{}_{\mu},P^{\nu_{1}\ldots\nu_{n}}{}_{\nu}\right ]=-n\delta_{\mu}^{(\nu_1|}P^{\mu_{1}\ldots\mu_{m}|\nu_2\ldots\nu_n)}{}_{\nu}+m\delta_{\nu}^{(\mu_1|}P^{\nu_{1}\ldots\nu_{n}|\mu_2\ldots\mu_m)}{}_{\mu}\ .\label{DiffCommutators}
\ee
The gauged $ISO(1,d)$ Poincar\'e algebra is generated by $\{P^{\mu_1\ldots\mu_m}{}_{a},J^{\nu_{1}\ldots\nu_{n}}{}_{ab}\}$, with commutation relations
\begin{align}
\left [P^{\nu_{1}\ldots\nu_{n}}{}_{a},P^{\mu_1\ldots\mu_m}{}_{b}\right ]&=0~,\nn
\left [P^{\nu_{1}\ldots\nu_{n}}{}_{a},J^{\mu_1\ldots\mu_m}{}_{bc}\right ]&=\eta_{ab}P^{\nu_{1}\ldots\nu_{n}\mu_1\ldots\mu_{m}}{}_{c}-\eta_{ac}P^{\nu_{1}\ldots\nu_{n}\mu_1\ldots\mu_{m}}{}_{b}~,\nn
 \left [J^{\nu_{1}\ldots\nu_{n}}{}_{ab},J^{\mu_1\ldots\mu_m}{}_{cd}\right ]&=-2\eta_{a[c}J^{\nu_{1}\ldots\nu_{n}\mu_1\ldots\mu_{m}}{}_{|b|d]}+2\eta_{b[c}J^{\nu_{1}\ldots\nu_{n}\mu_1\ldots\mu_{m}}{}_{|a|d]}\label{PoincareAlgebraLocal}\ .
\end{align} 
The two sets of generators do not commute and instead satisfy the following relations
\begin{align}
\left [P^{\alpha_{1}\ldots\alpha_{m}}{}_{\mu},P^{\nu_{1}\ldots\nu_{n}}{}_{a}\right ]&=-n\delta_{\mu}^{(\nu_{1}|}P^{\alpha_{1}\ldots\alpha_{m}|\nu_{2}\ldots\nu_{n})}{}_{a}~,\nn
\left [P^{\alpha_{1}\ldots\alpha_{m}}{}_{\mu},J^{\nu_{1}\ldots\nu_{n}}{}_{ab}\right ]&=-n\delta_{\mu}^{(\nu_{1}|}J^{\alpha_{1}\ldots\alpha_{m}|\nu_{2}\ldots\nu_{n})}{}_{ab}\ .\label{IntercommutingAlgebra}
\end{align}
We will need some notation to refer to these groups and their various subgroups.
We denote the local Poincar\'e group by $ISO(1,d)_{\rm local}$, its local subgroup generated by $J^{\mu_1\ldots\mu_n}{}_{ab}$, for all $n\geq 0$, by $SO(1,d)_{\rm local}$ and the global group generated by $J_{ab}$ as $SO(1,d)_{\rm global}$.  The diffeomorphism group is denoted by ${\rm Diff}(d+1)$ and the subgroup generated by $P^{\nu}{}_{\mu}$ -- which generates linear transformations -- is denoted by $GL(d+1)$.  Finally, there is a Lorentz subgroup of $GL(d+1)$ generated by $P_{[\mu\nu]}$, where $P_{\mu\nu}\equiv \eta_{\mu\sigma}P^{\sigma}{}_{\nu}$, which will be important later and which we denote by $SO(1,d)_{\rm spacetime}$.  In summary,
\begin{align} &\underset{P_{[\mu\nu]}}{SO(1,d)_{\rm spacetime}}\subset \underset{P^{\nu}{}_{\mu}}{GL(d+1)}\subset \underset{P^{\nu_1\ldots\nu_{n}}{}_{\mu}}{{\rm Diff}(d+1)}~,  \nn 
& \nn
& \underset{J_{ab}}{SO(1,d)_{\rm global}}\subset \underset{J^{\nu_{1}\ldots\nu_{n}}{}_{ab}}{SO(1,d)_{\rm local}}\subset  \underset{P^{\mu_1\ldots\mu_m}{}_{a},J^{\nu_{1}\ldots\nu_{n}}{}_{ab}}{ISO(1,d)_{\rm local}}~. 
\end{align}
In order to elucidate which symmetries should be realized linearly or nonlinearly, we recall how these symmetries act in the vielbein formalism. In these variables, the fields of GR are the vielbein $e_\mu^{\ a}$ and the spin connection $\omega_{\mu}{}^{ab}$, which is anti-symmetric in its Lorentz indices.  In the coset construction of Yang--Mills, the transformation properties of the connection (the gauge field) guide the choice of linearly realized subgroups, and we apply the same logic to GR.  The spin connection transforms under a local Lorentz transformation $\Lambda^a_{\ a'}(x)$ as
\be
\omega_{\mu}{}^{ab}\longmapsto \Lambda^{a}{}_{a'}\Lambda^{b}{}_{b'} \omega_{\mu}{}^{a'b'}-\Lambda^{bc}\partial_{\mu}\Lambda^{a}{}_{c}, 
\ee
and only global Lorentz transformations (for which $\partial\Lambda = 0$) are linearly realized.  Similarly, by examining the diffeomorphisms we see that the only linear transformations are those generated by the $GL(d+1)$ subgroup.  Hence, the breaking pattern we ought to consider is $G\to H$ with\footnote{Strictly speaking, $G$ and $H$ are not direct products due to the non-trivial commutation relations between the factors \eqref{IntercommutingAlgebra}, but we abuse notation slightly, with this caveat understood.}.  $G=ISO(1,d)_{\rm local}\times {\rm Diff}(d+1)$ and $H= SO(1,d)_{\rm global}\times GL(d+1)$.
    
\subsection{Calculation of the Maurer--Cartan form}
    
Now that we have a candidate symmetry breaking pattern, we can employ the coset machinery to construct building blocks that transform nicely under the symmetries. A representative element of $G/H$ is conveniently written as
\be
g\equiv e^{x^{\mu}P_{\mu}}e^{\phi_{\mu\nu}{}^{a}P^{\mu\nu}{}_{a}}e^{\phi_{\mu}{}^{a}P^{\mu}{}_{a}}e^{\phi^{a}P_{a}}e^{\frac{1}{2}\Theta_{\mu\nu}{}^{ab}J^{\mu\nu}{}_{ab}}e^{\frac{1}{2}\Theta_{\mu}{}^{ab}J^{\mu}{}_{ab}}\left (\cdots\right ) \ , 
\label{CosetElementDiffsAndLLTs}
\ee
where $(\cdots)$ contains only higher order factors involving fields along $P^{\mu\nu\rho}{}_{a}$, $J^{\mu\nu\rho}{}_{ab}$ or generators with even more Greek indices, which will not be relevant to our calculation.  For instance, components along generators of the form $P^{\nu\rho\ldots}{}_{\mu}$ do not enter our construction at all, though the role of the $GL(d+1)$ connection and related terms can be explored in Appendix \ref{Appendix:GLConnection}.

The Maurer--Cartan form is expanded as
\be
g^{-1}\rd g\equiv \Omega=\Omega^{\mu}P_{\mu}+\Omega^{a}P_{a}+\Omega_{\mu}{}^{a}P^{\mu}{}_{a}+\frac{1}{2}\Omega^{ab}J_{ab}+\frac{1}{2}\Omega_{\mu}{}^{ab}J^{\mu}{}_{ab}+\ldots\label{OmegaDiffLLTdefinition}
\ee
Explicit calculation using~\eqref{CosetElementDiffsAndLLTs}, the Baker--Campbell--Hausdorff formula, and the commutation relations~\eqref{DiffCommutators}--\eqref{IntercommutingAlgebra} gives
\begin{align}
\Omega^{\mu}&=\rd x^{\mu}~,\nn
\Omega^{a}&=\rd\phi^{a}-\rd x^{\mu}\phi_{\mu}{}^{a}~,\nn
\Omega_{\mu}{}^{a}&=\rd\phi_{\mu}{}^{a}-2\rd x^{\nu}\phi_{\nu\mu}{}^{a}+\rd\phi_{b}\Theta_{\mu}{}^{ba}-\rd x^{\nu}\phi_{\nu}{}^{b}\Theta_{\mu b}{}^{a}~,\nn
\Omega^{ab}&=-\rd x^{\mu}\Theta_{\mu}{}^{ab}~,\nn
\Omega_{\mu}{}^{ab}&=\rd\Theta_{\mu}{}^{ab}-2\rd x^{\nu}\Theta_{\nu\mu}{}^{ab}-\rd x^{\nu}\Theta_{\nu}{}^{[a|c|}\Theta_{\mu c}{}^{b]}~,\label{GRMCcomponents}
\end{align}
where Latin indices are raised and lowered with the constant flat metric $\eta_{ab}$.

\subsection{Identifications and inverse Higgs constraints}

There is an important identity satisfied by the components of the Maurer--Cartan form, known as the Maurer--Cartan equation. Consider a Lie algebra with generators $Q_{a}$, which obey the algebra $[Q_{a},Q_{b}]=f_{ab}^{~~c}Q_{c}$; the components of the associated algebra-valued Maurer--Cartan form defined through $\Omega\equiv\Omega^{a}Q_{a}$ are related by
\be
\rd\Omega^{a}=-\frac{1}{2}f_{bc}^{~~a}\Omega^{b}\wedge\Omega^{c}\ .
\ee
For the case at hand, this translates to the following identities satisfied by the MC form components \eqref{GRMCcomponents},
\begin{align}
\rd\Omega^{\mu}&=-\Omega_{\nu}{}^{\mu}\wedge\Omega^{\nu}~,\nn
\rd\Omega^{a}&=\Omega^{\mu}\wedge\Omega_{\mu}{}^{a}-\Omega^{b}\wedge \Omega_{b}{}^{a}~,\nn
\rd\Omega^{ab}&=\Omega^{\mu}\wedge\Omega_{\mu}{}^{ab}+\Omega^{ac}\wedge\Omega_{c}{}^{b}~,\nn
\rd\Omega_{\mu}{}^{a}&=2\Omega^{\nu}\wedge\Omega_{\mu\nu}{}^{a}-\Omega_{\mu}{}^{c}\wedge\Omega_{c}{}^{a}-\Omega^{c}\wedge\Omega_{\mu c}{}^{a}+\Omega_{\mu}{}^{\nu}\wedge\Omega_{\nu}{}^{a}~,\nn
\rd\Omega_{\mu}{}^{ab}&=2\Omega^{\nu}\wedge\Omega_{\nu\mu}{}^{ab}+2\Omega_{\mu}{}^{[a|c|}\wedge\Omega_{c}{}^{b]}+\Omega_{\mu}{}^{\nu}\wedge\Omega_{\nu}{}^{ab}~.\label{GRMCIdentities}
\end{align}
Given these relations between the various components of the MC form, we can infer their relation to the more familiar elements of differential geometry ${\bf e}^{a}$, $\boldsymbol\omega^{ab}$, ${\bf T}^{a}$ and ${\bf R}^{ab}$ (the vielbein 1-form, spin connection 1-form, torsion 2-form and curvature 2-form, respectively), which turn out to be the Cartan structure equations and Bianchi identities,
\begin{align}
\rd{\bf e}^{a}&= {\bf T}^{a}-{\bf e}^{b}\wedge\boldsymbol\omega_{b}{}^{a}~,\nn
\rd\boldsymbol\omega^{ab}&={\bf R}^{ab}-\boldsymbol\omega^{ac}\wedge\boldsymbol\omega_{c}{}^{b}~,\nn
\rd {\bf T}^{a}&={\bf T}^{c}\wedge\boldsymbol\omega_{c}{}^{a}-{\bf e}^{c}\wedge {\bf R}_{c}{}^{a}~,\nn
\rd {\bf R}^{ab}&={\bf R}^{ac}\wedge\boldsymbol\omega_{c}{}^{b}-{\bf R}^{bc}\wedge\boldsymbol\omega_{c}{}^{a}~,
\end{align}
leading us to tentatively identify 
\begin{align}
{\bf e}^{a}&=\Omega^{a}~,\nn
\boldsymbol\omega^{ab}&=\Omega^{ab}~,\nn
{\bf T}^{a}&=\Omega^{\mu}\wedge\Omega_{\mu}{}^{a}~,\nn
{\bf R}^{ab}&=\Omega^{\mu}\wedge\Omega_{\mu}{}^{ab}~ .
\label{GRidentifications}
\end{align}

The ``tentative" identification is due to the fact that the coset construction produces infinite towers of fields $\phi_{\mu_1\ldots\mu_m}{}^{a}$ and $\Theta_{\nu_{1}\ldots\nu_{n}}{}^{ab}$, while GR typically makes use of only the vielbein ${\bf e}^{a}$ and the spin connection $\boldsymbol\omega^{ab}$, with the latter independent or defined in terms of the vielbein depending on whether we work in the first or second order formulation.  In order to complete the reproduction of standard GR and eliminate the unnecessary fields we need to impose inverse Higgs constraints.  Referring to the rules for the inverse Higgs effect, we find that we can eliminate every field of the form $\phi_{\nu_{1}\ldots\nu_{n}}{}^{a}$ with $n\geq 1$ and all those of the  form $\Theta_{\mu_1\ldots\mu_m}{}^{ab}$ with $m\geq2$. At the order to which we are working, this simply means that $\phi_{\mu}{}^{a}$, $\phi_{\mu\nu}{}^{a}$ and $\Theta_{\mu\nu}{}^{ab}$ can potentially be removed. 

As is often the case, there is some art in choosing which fields to eliminate, and we will not want to remove all of them.  For instance, we will not want to eliminate $\phi_{\mu}{}^{a}$ as this would require setting ${\bf e}^{a}=0$, and to reproduce GR we cannot have a vanishing vielbein. It may be interesting to employ such a constraint in other contexts, but not in the present one. 

 Removing fields then requires setting parts of $\Omega_{\mu}{}^{a}$ and $\Omega_{\mu}{}^{ab}$ to zero, but we do not necessarily have to set \textit{all} components of these forms to zero.  More precisely, restoring form indices via $\Omega_{\mu}{}^{a}\equiv \Omega_{\nu\mu}{}^{a}\rd x^{\nu}$ and $\Omega_{\mu}{}^{ab}\equiv \Omega_{\nu\mu}{}^{ab}\rd x^{\nu}$, we find that $\phi_{\mu\nu}{}^{a}$ and $\Theta_{\mu\nu}{}^{ab}$ only appear in the symmetric components since they themselves are symmetric under $\mu\leftrightarrow\nu$,
\begin{align}
\Omega_{(\mu\nu)}{}^{a}&=\partial_{(\mu}\phi_{\nu)}{}^{a}-2\phi_{\mu\nu}{}^{a}+\partial_{(\mu}\phi_{|b|}\Theta_{\nu)}{}^{ba}-\phi_{(\mu}{}^{b}\Theta_{\nu)b}{}^{a}~,\nn
\Omega_{[\mu\nu]}{}^{a}&=\partial_{[\mu}\phi_{\nu]}{}^{a}+\partial_{[\mu}\phi_{|b|}\Theta_{\nu]}{}^{ba}-\phi_{[\mu}{}^{b}\Theta_{\nu]b}{}^{a}~,\nn
\Omega_{(\mu\nu)}{}^{ab}&=\partial_{(\mu}\Theta_{\nu)}{}^{ab}-2\Theta_{\mu\nu}{}^{ab}-\Theta_{(\mu}{}^{ac}\Theta_{\nu)c}{}^{b}~,\nn
\Omega_{[\mu\nu]}{}^{ab}&=\partial_{[\mu}\Theta_{\nu]}{}^{ab}-\Theta_{[\mu}{}^{ac}\Theta_{\nu]c}{}^{b}\ .
\end{align}
Therefore, if our goal is to eliminate only $\phi_{\mu\nu}{}^{a}$ and $\Theta_{\mu\nu}{}^{ab}$ then the appropriate inverse Higgs prescription is to demand that $\Omega_{(\mu\nu)}{}^{a}=0=\Omega_{(\mu\nu)}{}^{ab}$ and nothing more.  A one-form evaluated on the Inverse Higgs constraint will be denoted by $\Omega\big|_{\rm IH}$, and using these two conditions we find
\begin{align}
\Omega_{\mu}{}^{a}\big|_{\rm IH}&=\rd x^{\nu}\left [\partial_{[\nu}\phi_{\mu]}{}^{a}+\partial_{[\nu}\phi_{|b|}\Theta_{\mu]}{}^{ba}-\phi_{[\nu}{}^{b}\Theta_{\mu]b}{}^{a}\right ]~,\nn
\Omega_{\mu}{}	^{ab}\big|_{\rm IH}&=\rd x^{\nu}\left [\partial_{[\nu}\Theta_{\mu]}{}^{ab}-\Theta_{[\nu}{}^{ac}\Theta_{\mu]c}{}^{b}\right ]\ .
\end{align}
Rewritten in terms of the vielbein and spin connection, which have components
\begin{align}
e_{\mu}{}^{a}&=\partial_{\mu}\phi^{a}-\phi_{\mu}{}^{a}, \quad\quad \omega_{\mu}{}^{ab}=-\Theta_{\mu}{}^{ab}\ ,
\end{align}
these can be rewritten as
\begin{align}
\Omega_{\mu}{}^{a}\big|_{\rm IH}&=\rd x^{\nu}\left [-\partial_{[\nu} e_{\mu]}{}^{a}-e_{[\nu|b|}\omega_{\mu]}{}^{ba} \right ]~,\nn
\Omega_{\mu}{}	^{ab}\big|_{\rm IH}&=-\rd x^{\nu}\left [\partial_{[\nu}\omega_{\mu]}{}^{ab}+\omega_{[\nu}{}^{ac}\omega_{\mu]c}{}^{b}\right ]\ .
\end{align}
When combined with $\Omega^{\mu}$ using the wedge product, these turn into the usual expressions for the torsion and Riemann curvature 2-forms, respectively,
\begin{align}
\Omega^{\mu}\wedge\Omega_{\mu}{}^{a}&=\rd {\bf e}^{a}+{\bf e}_{b}\wedge\boldsymbol\omega^{ba}={\bf T}^{a}~,\nn
\Omega^{\mu}\wedge\Omega_{\mu}{}^{ab}&=\rd\boldsymbol\omega^{ab}+ \boldsymbol\omega^{ac}\wedge\boldsymbol\omega_{c}{}^{b}={\bf R}^{ab}.\label{TorsionAndRiemann}
\end{align}

Alternatively, we might desire to eliminate the field $\Theta_{\nu}{}^{ab}$ in addition to $\Theta_{\mu\nu}{}^{ab}$ and $\phi_{\mu\nu}{}^{a}$. This corresponds to removing all fields in favor of  $\phi^{a}$ and $\phi_{\mu}{}^{a}$, which are the only fields appearing in the definition of the vielbein \eqref{GRMCcomponents}, and hence we expect this to lead to zero torsion.  \textit{Both} the symmetric and antisymmetric parts of $\Omega_{\mu\nu}{}^{a}$ would then be set to zero, while only the symmetric part of $\Omega_{\mu\nu}{}^{ab}$ would vanish.\footnote{As pointed out in \cite{Delacretaz:2014oxa}, one could also choose inverse Higgs constraints to set all of $\Omega_{\mu\nu}{}^{ab}$ to zero which would lead to teleparallel gravity.}  This procedure leaves the identification $\Omega^{\mu}\wedge\Omega_{\mu}{}^{ab}={\bf R}^{ab}$ unchanged but now $\rd {\bf e}^{a}+{\bf e}_{b}\wedge\boldsymbol\omega^{ba}=0$; {\it i.e.}, the geometry would be torsion free.

  This is a standard ambiguity which arises in the formulation of gravity; in the first order formulation torsion is an independent variable, and is determined dynamically, so whether it vanishes or not depends on the form the action takes.  In the second order formulation its dependence on the vierbein, and hence the torsion, is prescribed.  For the remainder of the paper, we consider only cases in which the torsion vanishes, effectively committing to the second order formulation (except in the lowest order cases such as Einstein-Hilbert and dRGT where the two formulations are dynamically equivalent).
    
We should also note that the Riemann and torsion tensors are actually insensitive to whether or not some of the inverse Higgs constraints are imposed. That is, the contractions appearing in \eqref{TorsionAndRiemann} eliminate $\phi_{\mu\nu}{}^{a}$ and $\Theta_{\mu\nu}{}^{ab}$ automatically, and so whether or not we eliminate these two fields by inverse Higgs constraints is irrelevant to ${\bf R}^{ab}$ and ${\bf T}^{a}$. At the order to which we are working, the only inverse Higgs constraint that makes a qualitative difference is whether we want to eliminate $\Theta_{\nu}{}^{ab}$ as an independent field, i.e. set the torsion to zero, as discussed above.
  
\subsection{Transformation of the fields under the symmetries}
Now that we have computed the various components of the Maurer--Cartan form, we want to verify that they transform as they ought to under the nonlinearly realized symmetries.
  
First, we investigate how the broken $SO(1,d)_{\rm local}$ transformations are realized on  $g$ by using \eqref{MCtransformationcondition} with $g'\equiv \exp\left (\lambda_{\nu_1\ldots\nu_{n}}{}^{ab}J^{\nu_{1}\ldots\nu_{n}}{}_{ab}\right )$. The result is that the $\phi_{\nu_{1}\ldots\nu_{n}}{}^{a}$ and $\Theta_{\mu_1\ldots\mu_m}{}^{ab}$ fields shift under a local Lorentz transformation as
\begin{align}
\phi^{a}&\longmapsto\phi^{a}+\lambda_{(n)}^{a}{}_{b}(x)\phi^{b}~,\nn
\phi_{\mu}{}^{a}&\longmapsto\phi_{\mu}{}^{a}+\lambda_{(n)}^{a}{}_{b}(x)\phi_{\mu}{}^{b}+\partial_{\mu}\lambda_{(n)}^{a}{}_{b}(x)\phi^{b}~,\nn
\Theta_{\mu}{}^{ab}&\longmapsto\Theta_{\mu}{}^{ab}+\lambda_{(n)}^{a}{}_{c}(x)\Theta_{\mu}{}^{cb}+\lambda_{(n)}^{b}{}_{c}(x)\Theta_{\mu}{}^{ac}+\partial_{\mu}\lambda_{(n)}^{ab}(x) \ , 
\label{GRLLTtransformations}
\end{align}
where $\lambda^{ab}_{(n)}(x)\equiv x^{\nu_{1}}\ldots x^{\nu_{n}}\lambda_{\nu_{1}\ldots\nu_{n}}{}^{ab}$. Summing over all $n$, this implies that $\Omega^{a}$ and $\Omega^{ab}$ transform in the following way:
\begin{align}
\Omega^{a}&\longmapsto \Omega^{a}+\lambda^{a}{}_{b}(x)\Omega^{b}~,\nn
\Omega^{ab}&\longmapsto \Omega^{ab}+\lambda^{a}{}_{c}(x)\Omega^{cb}+\lambda^{b}{}_{c}(x)\Omega^{ac}-\rd\lambda^{ab}(x)\ ,
\end{align}
where $\lambda^a{}_b(x)$ is an arbitrary function.
These are simply the infinitesimal versions of
\begin{align}
\Omega^{a}&\longmapsto \Lambda^{a}{}_{b}\Omega^{b}~,\nn
\Omega^{ab}&\longmapsto \Lambda^{a}{}_{c}\Lambda^{b}{}_{d}\Omega^{cd}-\Lambda^{b}{}_{c}\rd\Lambda^{ac} \ ,	
\end{align}
with $\Lambda^{a}{}_{b}=\exp \left(\lambda^{a}{}_{b}\right)$.  Therefore, as expected, the nonlinearly realized $SO(1,d)_{\rm local}$ symmetries simply correspond to LLTs.

Next, we consider the transformation enacted through local internal translations, which are less familiar. That is, we calculate the transformation \eqref{MCtransformationcondition} with $g'=\exp\left ( c_{\nu_1\ldots\nu_n}{}^{a}P^{\nu_1\ldots\nu_n}{}_{a}\right )$ and find that the net effect is to change
\begin{align}
\phi_{\mu_1\ldots\mu_m}{}^{a}&\longmapsto \phi_{\mu_1\ldots\mu_m}{}^{a}+\frac{1}{m!}\partial_{\mu_1}\ldots\partial_{\mu_m}c_{(n)}^{a}(x)~,\nn
{\rm where}&~~~~~c_{(n)}^{a}(x)\equiv x^{\nu_1}\ldots x^{\nu_n}c_{\nu_1\ldots\nu_{n}}{}^{a}\label{LocalInternalTranslations}\ ,
\end{align}
 while the $\Theta_{\mu_1\ldots\mu_m}{}^{ab}$ fields are left unaffected. Again, summing over $n$ shows that there is a full function's worth of freedom.
It can be seen from \eqref{GRMCcomponents} that the Maurer-Cartan components are invariant under these shifts; all $c^{a}(x)$ factors cancel. Then, \eqref{LocalInternalTranslations} demonstrates that the $\phi^{a}$'s can be shifted around arbitrarily and so the internal local translations correspond to the ability to arbitrarily choose these coordinates of the coset space.   In particular, one can always use this freedom to go to a gauge in which the $\phi^{a}$'s coincide with the physical coordinates,  $\phi^{a}(x)=x^{\mu}$.

Finally, we explore diffeomorphisms by calculating the transformation \eqref{MCtransformationcondition} with  $\exp\left( c_{\alpha_{1}\ldots\alpha_{n}}{}^{\beta}P^{\alpha_{1}\ldots\alpha_{n}}{}_{\beta}\right)$.  Direct calculation yields the transformation rules
\begin{align}
x^{\mu}&\longmapsto x^{\mu}+c^{\mu}(x)~,\nn
\phi^{a}&\longmapsto\phi^{a}~,\nn
\phi_{\mu}{}^{a}&\longmapsto \left (\delta_{\mu }^{\mu'}-\partial_{\mu }c^{\mu'}(x)\right )\phi_{\mu'}{}^{a}~,\nn
\phi_{\mu\nu}{}^{a}&\longmapsto \left (\delta_{\mu }^{\mu'}-\partial_{\mu }c^{\mu'}(x)\right )\left (\delta_{\nu}^{\nu'}-\partial_{\nu}c^{\nu'}(x)\right )\phi_{\mu'\nu'}{}^{a}-\frac{1}{2}\phi_{\beta}{}^{a}\partial_{\mu}\partial_{\nu}c^{\beta}(x)~,\nn
\Theta_{\mu}{}^{ab}&\longmapsto \left (\delta_{\mu }^{\mu'}-\partial_{\mu }c^{\mu'}(x)\right )\Theta_{\mu'}{}^{ab}~,\nn
\Theta_{\mu\nu}{}^{ab}&\longmapsto \left (\delta_{\mu }^{\mu'}-\partial_{\mu }c^{\mu'}(x)\right )\left (\delta_{\nu}^{\nu'}-\partial_{\nu}c^{\nu'}(x)\right )\Theta_{\mu'\nu'}{}^{ab}-\frac{1}{2}\Theta_{\beta}{}^{ab}\partial_{\mu}\partial_{\nu}c^{\beta}(x)~.\label{GRFieldTransformationDiffs}
\end{align}
The effect of these transformations on the MC components is
\begin{align}
\Omega^{\mu}&\longmapsto\frac{\partial x'^{\mu }}{\partial x^{\nu}}\Omega^{\nu}~,\nn 
\Omega^{a}&\longmapsto\Omega^{a}~,\nn
\Omega_{\mu}{}^{a}&\longmapsto\frac{\partial x^{\nu}}{\partial x'^{\mu}}\Omega_{\nu}{}^{a}~,\nn
\Omega^{ab}&\longmapsto\Omega^{ab}~,\nn
\Omega_{\mu}{}^{ab}&\longmapsto\frac{\partial x^{\nu}}{\partial x'^{\mu}}\Omega_{\nu}{}^{ab}~,\label{GRDiffsLLTsMCComponents}
\end{align}
where $x'^{\mu}\equiv x^{\mu}+c^{\mu}(x)$. Therefore we have accurately reproduced diffeomorphisms, and for our actions to be invariant under these nonlinearly realized symmetries every upper Greek index on a MC form component needs to be contracted with a lower Greek index and vice versa.

Note the importance of including diffeomorphisms into the coset procedure from the beginning.  Had we not included them, there would be no transformation of Greek indices induced by left multiplication by a general group element as in~\eqref{MCtransformationcondition}.  Certainly, we could impose diffeomorphism invariance on such a construction by fiat, but this would not naturally lend itself to our later exploration of Higgs phases of GR where we wish to break the diffeomorphism symmetry, and so we find it more natural to include it in the coset from the start.

\subsection{Constructing the action for Einstein gravity}  
  
 We now have the standard ingredients needed to construct $(d+1)$-dimensional General Relativity.  The $SO(1,d)_{\rm global}\times GL(d+1)$ preserved symmetry informs us that we can build actions out of any component of the Maurer--Cartan form which does not lie along $J_{ab}$ or $P^{\mu}{}_{\nu}$ (which are linearly realized) and that we must contract all Latin indices with either $\eta_{ab}$ or $\epsilon_{a_{0}\ldots a_{d}}$ (in order to ensure local Lorentz invariance) while upper Greek indices must be contracted with lower Greek indices (to ensure diffeomorphism invariance).
 
   Following these rules, we can form the Einstein-Hilbert action as
\be
S_{\rm EH}\equiv \frac{1}{(d-1)!}\int \epsilon_{a_{0}\ldots a_{d}}\Omega^{a_{0}}\wedge\cdots\wedge\Omega^{a_{d-2}}\wedge\Omega^{\mu}\wedge\Omega_{\mu}{}^{a_{d-1}a_{d}}~,
\ee
which can be put in the more familiar form by using the identifications~\eqref{GRidentifications}
\be
 S_{\rm EH} = \frac{1}{(d-1)!}\int \epsilon_{a_{0}\ldots a_{d}}{\bf e}^{a_0}\wedge\cdots\wedge{\bf e}^{a_{d-2}}\wedge {\bf R}^{a_{d-1}a_d}=\int\rd^{d+1}x\sqrt{-g}R~.\label{GREH}
\ee
Similarly, the cosmological constant term can be constructed as
\be
S_{\Lambda}\equiv \frac{1}{(d+1)!} \int \Lambda~\epsilon_{a_{0}\ldots a_{d}}\Omega^{a_0}\wedge\cdots\wedge\Omega^{a_{d}} =\frac{1}{(d+1)!} \int \Lambda~\epsilon_{a_{0}\ldots a_{d}}{\bf e}^{a_0}\wedge\cdots\wedge{\bf e}^{a_0}=\int\rd^{d+1}x\sqrt{-g}\,\Lambda\ .\label{GREHandCC}
\ee

These two operators represent the lowest order terms in the sense that all other interactions will be at least quadratic in the set $\{\Omega_{\mu}{}^{a},\Omega_{\mu}{}^{ab}\}$. (Equivalently, all further interactions will involve more derivatives.)

 \subsubsection{Lovelock invariants}
The identities~\eqref{GRMCIdentities} also allow us to demonstrate the existence and topological nature of the Euler class in even dimensions.  For instance, in two dimensions we have ${\cal E}_2 = \frac{1}{8\pi}\epsilon_{ab}\Omega^{\mu}\wedge\Omega_{\mu}{}^{ab} = \frac{1}{4\pi} \rd^{2}x\sqrt{-g}R$, and the identities~\eqref{GRMCIdentities} tell us
\begin{align}
\rd\left (\epsilon_{ab}\Omega^{\mu}\wedge\Omega_{\mu}{}^{ab}\right )
&=-4\epsilon_{ab}\Omega^{\mu}\wedge\Omega_{\mu}{}^{ac}\wedge\Omega_{c}{}^{b}=0\ ,\label{GB2D}
\end{align}
since the Latin indices can only take on two values. This shows that the Ricci scalar is a total derivative in two dimensions.  This readily generalizes to the case of $d+1=2n$ dimensions where one can show that the following $2n$-form is closed
\be
{\cal E}_{2n}\equiv \frac{1}{(4\pi)^n n!}\epsilon_{a_{1}\ldots a_{2n}}\Omega^{\mu}\wedge\Omega_{\mu}{}^{a_1a_{2}}\wedge\ldots\wedge\Omega^{\mu}\wedge\Omega_{\mu}{}^{a_{2n-1}a_{2n}}~.
\label{lovelock}
\ee
In fact, this expression is nothing more than the $2n$-dimensional Euler density (also sometimes called the Lovelock invariants\footnote{These terms are distinguished in that they are the unique terms which lead to second order equations of motion for the metric in the absence of torsion~\cite{Lovelock:1971yv}.}), which can be seen by using~\eqref{GRidentifications} to write~\eqref{lovelock} as
\begin{align}
\nonumber
{\cal E}_{2n} &\equiv \frac{1}{(4\pi)^n n!}\epsilon_{a_{1}\ldots a_{2n}}{\bf R}^{a_1 a_2}\wedge\cdots\wedge {\bf R}^{a_{2n-1}a_{2n}} \\
&=  \frac{1}{(8\pi)^n n!}\rd^{2n}x\sqrt{-g}~\epsilon^{\mu_{1}\ldots \mu_{2n}}\epsilon^{\nu_{1}\ldots \nu_{2n}}R_{\mu_1\mu_2\nu_1\nu_2}\cdots R_{\mu_{2n-1}\mu_{2n}\nu_{2n-1}\nu_{2n}}~.
\end{align}
The Lovelock invariants can also be written away from their home dimension, but in this case they are no longer topological, but nevertheless retain second-order equations of motion. The above viewpoint on the Lovelock invariants may seem somewhat foreign, but it is simply a rephrasing of standard calculations, since the identities~\eqref{GRMCIdentities} are really just the Bianchi identities.

\subsubsection{Other topological invariants}
In addition to the Euler densities, there exist other topological invariants in GR. Most analogous to the Yang--Mills case, there exist Pontryagin classes in $(d+1) = 4k$ dimensions, which take the form
\be
{\cal P}_{4k} = {\rm tr}~\Omega^{\mu_1}\wedge\Omega_{\mu_1}^{a_1 b_1}\wedge\cdots\wedge \Omega^{\mu_{2k}}\wedge\Omega_{\mu_{2k}}^{a_{2k} b_{2k}} = {\rm tr}~{\bf R}^{a_1 b_1}\wedge\cdots\wedge{\bf R}^{a_{2k} b_{2k}}~.
\ee
Using the identities~\eqref{GRMCIdentities}, it can be checked that these forms are closed. Notice that, as in the Yang--Mills case, in higher dimensions, there are many inequivalent Pontryagin terms. For example in $(d+1) = 8$, both ${\cal P}_8$ and ${\cal P}_4\wedge{\cal P}_4$ can be present, and associated to each of the Pontryagin terms is a Chern--Simons term (see~\cite{Goon:2014ika} for details of how this appears in our language, the situation is essentially identical to that of Yang--Mills.) Finally we note in passing that, in the presence of non-zero torsion, there is an additional characteristic class called the Nieh--Yan class, see~\cite{Mardones:1990qc,Zanelli:2012px} for more details.
 
\section{Massive gravity}
\label{massivegconstruct}
Having employed coset machinery to build Einstein gravity, we now turn to the construction of massive gravity by analogous methods. In order to think of massive gravity as a Higgs-ed phase of Einstein gravity, it should linearly realize fewer symmetries. Physically, some unknown mechanism will reduce the preserved symmetry group from $SO(1,d)_{\rm global}\times GL(d+1)$ to some smaller subgroup, nonlinearly realizing more symmetries. Since having a smaller preserved subgroup allows for a larger number of terms in the action, we are looking for the largest preserved group which will still admit the dRGT mass terms. Ideally, there would exist a symmetry breaking pattern such that the dRGT terms provide the only possible interactions, but this is probably too optimistic because explicit computations indicate that interactions which are not of the dRGT form are generated quantum-mechanically~\cite{deRham:2013qqa}.\footnote{The situation can be different when additional degrees of freedom are present}
 \subsection{Symmetries and algebras}
 Inspecting the dRGT action \eqref{dRGTMassTerms}, we see the important role played by the unit one form ${\bf 1}^{a}=\delta_{\mu}^{a}\rd x^{\mu}$.  The $\delta_{\mu}^{a}$ tensor is not invariant under either $SO(1,d)_{\rm global}$ or $GL(d+1)$ independently, but {\it is} invariant under the diagonal combination of $SO(1,d)_{\rm global}$ and the subgroup $SO(1,d)_{\rm spacetime}\subset GL(d+1)$ whose generators generators can be taken to be $P_{[\mu\nu]}+\frac{1}{2}\delta^a_\mu \delta^b_\nu J_{ab}$.  (Since the new invariant tensor allows us to freely change between Greek and Latin indices we will ignore this distinction when convenient.)
We thus posit that the appropriate breaking pattern to study is 
\be 
ISO(1,d)_{\rm local}\times {\rm Diff}(d+1) \longrightarrow \left(SO(1,d)_{\rm global}\times SO(1,d)_{\rm spacetime}\right )_{\rm diag}~.
\ee
In addition to the broken generators of the GR case, the generators $P_{(\mu\nu)}$ are now broken, along with a combination of $P_{[\mu\nu]}$ and $\frac{1}{2}J_{ab}$ linearly independent of $P_{[\mu\nu]}+\frac{1}{2}J_{ab}$, which we take to be simply $\frac{1}{2}J_{ab}$. 
 
\subsubsection{$\left(SO(1,d)_{\rm global}\times SO(1,d)_{\rm spacetime}\right )_{\rm diag}$ as the optimal subgroup}
 
It is possible to demonstrate that $H = \left(SO(1,d)_{\rm global}\times SO(1,d)_{\rm spacetime}\right )_{\rm diag}$ is the largest symmetry subgroup we can preserve which will still admit the dRGT interactions.  Consider a generic dRGT interaction, say
\be
 \mathcal{L}\sim \epsilon_{abcd}{\bf 1}^{a}\wedge{\bf 1}^{b}\wedge {\bf e}^{c}\wedge{\bf e}^{d}\sim \rd^{4}x\, \epsilon^{\mu\nu\rho\sigma}\epsilon_{abcd}\delta_{\mu}^{a}\delta_{\nu}^{b}e_{\rho}{}^{c}e_{\sigma}{}^{d}  \ .\label{exampledRGTinteraction}
\ee
Under an infinitesimal gauged Lorentz transformation the vielbeins transform as
\be
 e_{\mu}{}^{a}\longmapsto \Big(\delta^{a}_{b}+\lambda^{a}{}_{b}(x)\Big)e_{\mu}{}^{b}\ ,
\ee
with $\lambda_{ab}(x)=\eta_{aa'}\lambda^{a'}{}_{b}(x)$ antisymmetric under $a\leftrightarrow b$, and under an infinitesimal diffeomorphism the vielbeins transform as
\be
e_{\mu}{}^{a}\longmapsto \Big(\delta^{\nu}_{\mu}-\partial_{\mu}c^{\nu}(x)\Big)e_{\nu}{}^{	a}\ .
\ee
 Given the infinitesimal Lorentz transformation, the interaction \eqref{exampledRGTinteraction} will only be invariant if we can choose $c^{\mu}(x)$ such that $\partial_{\nu}c^{\mu}(x)=\lambda^{\mu}{}_{\nu}$.  Expanding $c^{\mu}(x)$ in a power series as
\be
 c^{\mu}(x)=\sum_{n}\frac{1}{n!}c^{\mu}{}_{\nu_1\ldots\nu_{n}}x^{\nu_{1}}\ldots x^{\nu_{n}} \ ,
\ee
where $c^{\mu}{}_{\nu_1\ldots\nu_{n}}$ is symmetric in its $\nu$ indices, we see that if $\lambda^{\mu}{}_{\nu}$ is a global transformation ({\it i.e.}, independent of $x$) then we can satisfy our condition by taking $c^{\mu}(x)=c^{\mu}{}_{\nu}x^{\nu}$, with $c^{\mu}{}_{\nu}=\lambda^{\mu}{}_{\nu}$. 
 
However, if the Lorentz transformation is $x$-dependent, we will need non-trivial $c^{\mu}{}_{\nu_1\ldots\nu_{n}}$ coefficients for $n>1$.  Lowering the $\mu$ index with $\eta$, we find we need to satisfy $\partial_{\nu}c_{\mu}(x)=\lambda_{\mu\nu}$ and hence $c_{\mu\nu_{1}\ldots\nu_{n}}$ must be antisymmetric under $\mu\leftrightarrow \nu_{i}$.  Using this fact and the symmetry in the $\nu$ indices, multiple permutations of the indices $\{\mu,\nu_1,\nu_2\}$ lead to $c_{\mu\nu_1\nu_{2}\ldots\nu_{n}}=-c_{\mu\nu_{1}\nu_{2}\ldots\nu_{n}}$ and hence we cannot satisfy our condition for $n>1$.  Therefore, we can preserve the global Lorentz symmetry while retaining the dRGT interactions, but none of the local Lorentz transformations can remain.
  
Had we started by performing a diffeomorphism and attempted to compensate with LLT's, a similar procedure would result and we would again be lead to the present diagonal subgroup which is therefore the optimal one for the present study.

 \subsection{Computing the Maurer--Cartan form}
 
Having now specified the symmetry breaking pattern, we can compute the Maurer--Cartan form.  This computation will closely mirror the procedure for Einstein gravity, with the exception that the new coset $G/H$ contains a few more broken generators than were included in the GR case, namely $P_{(\mu\nu)}$ is now broken, along with $J_{ab}$. The massive gravity coset element will then contain two more factors than the GR representative element: $e^{\psi^{\mu}_{\nu}P_{\mu}^{\nu}}e^{\frac{1}{2}\Theta^{ab}J_{ab}}$, where $\psi^{\mu}_{\nu}$ is symmetric.  That is, we have
\begin{align}
g_{\rm mg}&\equiv e^{x^{\mu}P_{\mu}}e^{\phi_{\mu\nu}{}^{a}P^{\mu\nu}{}_{a}}e^{\phi_{\mu}{}^{a}P^{\mu}{}_{a}}e^{\phi^{a}P_{a}}e^{\frac{1}{2}\Theta_{\mu\nu}{}^{ab}J^{\mu\nu}{}_{ab}}e^{\frac{1}{2}\Theta_{\mu}{}^{ab}J^{\mu}{}_{ab}}\left (\cdots\right )e^{\psi^{\mu}_{\nu}P_{\mu}^{\nu}}e^{\frac{1}{2}\Theta^{ab}J_{ab}}\nn
&= g_{\rm GR}e^{\psi^{\mu}_{\nu}P_{\mu}^{\nu}}e^{\frac{1}{2}\Theta^{ab}J_{ab}} \ ,
\label{TypicalCosetElementdRGTcase}
\end{align}
where $g_{\rm GR}$ is the representative coset element in \eqref{CosetElementDiffsAndLLTs}.

The Maurer--Cartan form is expanded as
\begin{align}
g_{\rm mg}^{-1}\rd g_{\rm mg}&\equiv \Omega=\Omega^{\mu}P_{\mu}+\Omega^{a}P_{a}+\Omega_{\mu}{}^{a}P^{\mu}{}_{a} +\frac{1}{2}\Omega^{ab}J_{ab}+\frac{1}{2}\Omega_{\mu}{}^{ab}J^{\mu}{}_{ab}+\ldots
\end{align}
and we only work to the order indicated.  Because the representative element at hand is so closely related to the one in the GR case, it does not take much extra work to calculate the Maurer--Cartan components.  The result is
\begin{align}
\Omega^{\mu}&=\Psi_{\nu}{}^{\mu}\rd x^{\nu}~,\nn
\Omega^{a}&=\tilde\Theta_{b}{}^{a}\left (\rd \phi^{b}-\rd x^{\mu}\phi_{\mu}{}^{b}\right )~,\nn
\Omega_{\mu}{}^{a}&=\tilde\Theta_{b}{}^{a}(\Psi^{-1})_{\mu}{}^{\nu}\big (\rd\phi_{\nu}{}^{b}-2\rd x^{\rho}\phi_{\rho\nu}{}^{b}+\rd\phi_{c}\Theta_{\nu}{}^{cb}-\rd x^{\rho}\phi_{\rho}{}^{c}\Theta_{\nu c}{}^{b}\big )~,\nn
\Omega^{ab}&=\tilde\Theta_{c}{}^{a}\tilde\Theta_{d}{}^{b}\left (-\rd x^{\mu}\Theta_{\mu}{}^{cd}\right )+\tilde\Theta^{ac}\rd \tilde\Theta^{b}{}_{c}~,\nn
\Omega_{\mu}{}^{ab}&=(\Psi^{-1})_{\mu}{}^{\nu}\tilde\Theta_{c}{}^{a}\tilde\Theta_{d}{}^{b}\big (\rd\Theta_{\nu}{}^{cd}-2\rd x^{\rho}\Theta_{\rho\nu}{}^{cd}-\rd x^{\rho}\Theta_{\rho}{}^{ce}\Theta_{\nu e}{}^{d}\big )~,\label{dRGTMCComponentsStuckelberged}
\end{align}
where we have defined $\Psi_{\mu}{}^\nu \equiv e^{-\psi_{\mu}{}^{\nu}}\equiv \delta_{\mu}{}^{\nu}-\psi_{\mu}{}^{\nu}+\frac{1}{2}\psi_{\mu}{}^{\alpha}\psi_{\alpha}{}^{\nu}-\ldots$ and similarly $\tilde\Theta_{c}{}^{a}=e^{\Theta_{c}{}^{a}}$.  We see that the $\tilde\Theta_{a}{}^{b}$ fields simply generate an LLT on the Latin indices, while the upper Greek indices get contracted with $\Psi_{\mu}{}^{\nu}$ and lower Greek indices are contracted by the inverse, $(\Psi^{-1})_{\mu}{}^{\nu}$ .

\subsection{Identifications and inverse Higgs constraints}

Because the new factors involving $\Theta^{ab}$ and $\psi_{\mu}{}^{\nu}$ only appear as overall multiplicative factors, we can impose the same inverse Higgs constraints as before.  Having done so, we identify
\begin{align}
\Omega^{a}&=\tilde\Theta_{b}{}^{a}{\bf e}^{b}~,\nn
\Omega^{\mu}\wedge\Omega_{\mu}{}^{a}&=\tilde\Theta_{b}{}^{a}{\bf T}^{b}~,\nn
\Omega^{\mu}\wedge\Omega_{\mu}{}^{ab}&=\tilde\Theta_{c}{}^{a}\tilde\Theta_{d}{}^{b}{\bf R}^{cd}\ .
\end{align}

\subsection{Symmetry transformations}

We now explore the nonlinearly realized symmetries in this new setup.  In the GR case the preserved subgroup was a product, $SO(1,d)_{\rm global}\times GL(d+1)$, and so nonlinearly realized symmetries could act on the components of the MC form as an element of either factor of this group, as we saw for LLT's and diffeomorphisms respectively.  Here, the preserved subgroup is no longer a product and so the transformations which once acted as LLT's or diffeomorphisms will instead act as an element of $\left (SO(1,d)_{\rm global}\times SO(1,d)_{\rm spacetime}\right )$.

In order to get a sense for what to expect for the transformation rules of the MC components we inspect
\begin{align}
\Omega_{\mu}{}^{a}&=\tilde\Theta_{b}{}^{a}(\Psi^{-1})_{\mu}{}^{\nu}\big (\rd\phi_{\nu}{}^{b}-2\rd x^{\rho}\phi_{\rho\nu}{}^{b}+\rd\phi_{c}\Theta_{\nu}{}^{cb}-\rd x^{\rho}\phi_{\rho}{}^{c}\Theta_{\nu c}{}^{b}\big )\nn
&=\tilde\Theta_{b}{}^{a}(\Psi^{-1})_{\mu}{}^{\nu}\Omega_{{\rm GR}\nu}{}^{b} ~,
\end{align}
where $\Omega_{{\rm GR}\nu}{}^{b}$ is the MC component along $P^{\nu}{}_{b}$ in the GR calculation. We first consider the transformations which in the GR case led to LLT's and follow with a study of those which led to diffeomorphisms.

We know that under the transformation \eqref{MCtransformationcondition} by an element generated by $J^{\nu_1\ldots\nu_n}{}_{ab}$, $\Omega_{{\rm GR}\mu}{}^{a}$ transforms by an LLT, i.e. $\Omega_{{\rm GR}\mu}{}^{b}\mapsto \Lambda^{b}{}_{c}\Omega_{{\rm GR}\mu}{}^{c}$.  Since there does not exist any series of commutators involving $J^{\nu_1\ldots\nu_n}{}_{ab}$ that can generate an element along $P^{\mu}{}_{\nu}$, the $\psi_{\mu}{}^{\nu}$ field (and thus $\Psi$) must stay invariant and only the $\tilde\Theta_{b}{}^{a}$ field can transform.  The $\Omega_{\mu}{}^{a}$ one-form must change by an element of the full diagonal group, meaning {both} the Greek and Latin indices must transform simultaneously.  Since we know the Greek indices do not change, neither can the Latin ones, and we must have that $\tilde\Theta_{b}{}^{a}$ transforms as 
\begin{align}
\tilde\Theta_{b}{}^{a}\longmapsto (\Lambda^{-1})^{c}{}_{b}\tilde\Theta_{c}{}^{a}\ .\label{LLTTTildeTransformation}
\end{align}
  Explicit calculation bears this out, and the set $\{\Omega^{\mu},\Omega^{a},\Omega_{\mu}{}^{a},\Omega_{\mu}{}^{ab}\}$ is left invariant by these transformations.  
 
Moving on, we examine the transformation \eqref{MCtransformationcondition} generated by an element $P^{\nu_{1}\ldots\nu_{n}}{}_{\mu}$ which we know causes $\Omega_{{\rm GR}\mu}{}^{a}$ to change as $\Omega_{{\rm GR}\mu}{}^{a}\mapsto \frac{\partial x^{\nu}}{\partial x'^{\mu}}\Omega_{{\rm GR}\nu}{}^{a}$.  In this case, there \textit{are} series of commutators involving $P^{\nu_{1}\ldots\nu_{n}}{}_{\mu}$ that have elements along both $J_{ab}$ and $P_{[\mu\nu]}$ and so both $\tilde\Theta_{a}{}^{b}$ and $\Psi_{\mu}{}^{\nu}$ can transform.  Then, since the overall MC form has to transform by an element of the diagonal group, we must have
 \begin{align}
\Psi_{\mu}{}^{\nu}&\longmapsto \Lambda(x')_{\rho}{}^{\nu} \Psi_{\sigma}{}^{\rho}\frac{\partial x'^{\sigma}}{\partial x^{\mu}}~,\nn
\tilde\Theta_{b}{}^{a}&\longmapsto \Lambda(x')_{c}{}^{a} \tilde\Theta_{b}{}^{c}\ .\label{DiffPsiandTTildeTransformation}
 \end{align}
 That is, the $\Psi_{\mu}{}^{\nu}$ field must absorb the diffeomorphism transformation on the right, since these are no longer a symmetry, and then both fields must transform on the left by the \textit{same} Lorentz transformation, $\Lambda(x')$, which depends on the relation between $x$  and $x'$ and enforces that $\Lambda(x')_{\rho}{}^{\nu} \Psi_{\sigma}{}^{\rho}\frac{\partial x'^{\sigma}}{\partial x^{\mu}}$ remain symmetric when $\mu$ and $\nu$ are lowered.  Again, explicit calculation bears this out, and these are the correct transformation laws.  

Note that the $\tilde\Theta_{b}{}^{a}$ and $\Psi_{\mu}{}^{\nu}$ fields are analogous to the St\"uckelberg fields that are more familiar in treatments of massive gravity \cite{ArkaniHamed:2002sp} (see \cite{Gabadadze:2013ria,Ondo:2013wka} for St\"uckelbergs in the vielbein formalism). These fields restore the $ISO(1,d)_{\rm local}\times {\rm Diff}(d+1)$ invariance of the original theory in much the same way. However, they differ from conventional St\"uckelberg fields in that they do not appear derivatively in the action.

Another important difference between the fields $\{\tilde{\Theta}_{b}{}^{a},\Psi_{\mu}{}^{\nu}\}$ and more conventional St\"uckelberg fields is the inability for us to go to the standard unitary gauge.  There are only $(d+2)(d+1)/2$ gauge transformation available in the preserved symmetry group, whereas the St\"uckelberg fields contain $(d+1)^{2}$ total components, and hence one cannot set a gauge in which $\tilde{\Theta}_{b}{}^{a}=\delta_{b}^{a}$ and $\Psi_{\mu}{}^{\nu}=\delta_{\mu}^{\nu}$ \textit{simultaneously}, which would be the analogue of the usual unitary gauge choice for demonstrating the dynamical equivalence of a theory in its St\"uckelberged and non-St\"uckelberged forms.    We return to this point in the next section when we construct a theory of massive gravity in the present language and argue the equivalence to dRGT.

\subsection{Constructing the action in massive gravity}

With fewer symmetries preserved, there are a greater number of allowed invariant contractions.  In particular, the fact that both Latin and Greek indices now change only by Lorentz transformations and that they must further transform by the \textit{same} Lorentz transformation implies that we can now use the $\eta$ and $\epsilon$ tensors to perform contractions and these can have any combination of Greek and Latin indices.  As long as all indices are contracted, the result will be invariant under the diagonal combination of Lorentz transformations.

\subsubsection{dRGT terms}

Since the GR operators we wrote down are invariant under a larger symmetry group that contains the one presently under study, we can still form all of the same operators that we had in the GR case.  Also, since these are fully $SO(1,d)_{\rm spacetime}\times SO(1,d)_{\rm global}$ invariant they are independent of the $\tilde\Theta_{b}{}^{a}$ and $\Psi_{\mu}{}^{\nu}$ fields, which is expected since these fields are responsible for restoring these symmetries. For instance, the Einstein--Hilbert and cosmological constant pieces are essentially identical to~\eqref{GREH} and~\eqref{GREHandCC}\ ,
 \begin{align}
\mathcal{L}_{R}&\equiv \frac{1}{(d-1)!}\int\epsilon_{a_{0}\ldots a_{d}}\Omega^{a_{0}}\wedge\ldots\wedge\Omega^{a_{d-2}}\wedge\Omega^{\mu}\wedge\Omega_{\mu}{}^{a_{d-1}a_{d}}=\int\rd^{d+1}x\sqrt{-g}R~,\nn
\mathcal{L}_{\Lambda}&\equiv \frac{1}{(d+1)!} \int \Lambda\epsilon_{a_{0}\ldots a_{d}}\Omega^{a_0}\wedge\ldots\wedge\Omega^{a_{d}}= \int\rd^{d+1}x\sqrt{-g}\,\Lambda\ .\label{dRGTEHandCC}
\end{align}	

However, there now exist new operators we can write down in the broken phase that we could not before. For concreteness, we specialize to $d+1=4$ for the remainder of this section, but nothing we say is really dependent on that choice.  We again set the torsion to zero, so the ingredients at our disposal to construct actions are $\{\Omega^{\mu},\Omega^{a},\Omega^{ab},\Omega_{\mu}{}^{ab}\}$; only $\Omega^{\mu}$ and $\Omega^{a}$ are free from derivatives.  Therefore, the lowest order terms in the derivative expansion will be of the form $\mathcal{L}\left (\Omega^{\mu},\Omega^{a}\right )$ and some of the simplest terms we can write are\footnote{Note that we could also construct terms with 4 factors of either $\Omega^\mu$ or $\Omega^a$, but the first will be a constant, and the latter corresponds to the cosmological constant considered above.}
\begin{align}
\mathcal{L}_{1}& = \epsilon_{\mu bcd}\Omega^{\mu}\wedge\Omega^{b}\wedge\Omega^{c}\wedge\Omega^{d} = \epsilon_{abcd}\Psi_{\mu}{}^a \rd x^{\mu} \wedge \tilde\Theta_{i}{}^{b}{\bf e}^{i}\wedge\tilde\Theta_{j}{}^{c}{\bf e}^{j}\wedge\tilde\Theta_{k}{}^{d}{\bf e}^{k} ~,\nn
\mathcal{L}_{2}& =  \epsilon_{\mu\nu cd}\Omega^{\mu}\wedge\Omega^{\nu}\wedge\Omega^{c}\wedge\Omega^{d} = \epsilon_{abcd}\Psi_{\mu}{}^a \rd x^{\mu} \wedge \Psi_{\nu}{}^b\rd x^{\nu}\wedge\tilde\Theta_{j}{}^{c}{\bf e}^{j}\wedge\tilde\Theta_{k}{}^{d}{\bf e}^{k}~,\nn
\mathcal{L}_{3}& = \epsilon_{\mu\nu\rho d}\Omega^{\mu}\wedge\Omega^{\nu}\wedge\Omega^{\rho}\wedge\Omega^{d}=\epsilon_{abcd}\Psi_{\mu}{}^a \rd x^{\mu} \wedge \Psi_{\nu}{}^b\rd x^{\nu}\wedge\Psi_{\rho}{}^c\rd x^{\rho}\wedge\tilde\Theta_{k}{}^{d}{\bf e}^{k}\ .\label{dRGTSectionMassTerms}
\end{align}

\subsubsection{Dynamical equivalence}

The Lagrangians in the previous section are clearly reminiscent of those of dRGT, but with the inclusion of factors of $\tilde\Theta_{b}{}^{a}$ and $\Psi_{\mu}{}^{\nu}$.  We now return to the discussion of the dynamical equivalence and argue that the theory defined by operators in  \eqref{dRGTEHandCC} and \eqref{dRGTSectionMassTerms} is equivalent to the usual dRGT theory.  As mentioned previously, we are unable to fix a gauge in which $\Psi_{\mu}{}^{a}=\delta_{\mu}^{a}$ and $\tilde{\Theta}_{b}{}^{a}=\delta_{b}^{a}$ simultaneously, which would constitute a manifest proof, but we will still be able to argue that the equivalence holds.

We start by fixing only the latter condition listed above, $\tilde{\Theta}_{b}{}^{a}=\delta_{b}^{a}$, which can always be achieved via the residual transformations in \eqref{LLTTTildeTransformation} and \eqref{DiffPsiandTTildeTransformation}.  This choice ensures that $\Omega^{a}$ coincides with the vielbein of the previous sections, $\Omega^{a}={\bf e}^{a}$.  Six out of the ten gauge degrees of freedom have been used, leaving us with the ability to only impose four more conditions on the $\Psi_{\mu}{}^{\nu}$ field. We note that in the parametrization $\Psi_{\mu}{}^{\nu}=e^{\psi_{\mu}{}^{\nu}}=\delta_{\mu}^{\nu}+\psi_{\mu}{}^{\nu}+\ldots$ the field $\psi_{\mu}{}^{\nu}$ appears in \eqref{dRGTSectionMassTerms} linearly in only one of four different possible tensor contractions.  Explicitly, we can write
\be
\sum_{i}\beta_{i}\mathcal{L}_{i}=f_{ 0 }(\beta_{i},{\bf e}){\rm Tr}[  \psi]+f_{ 1 }(\beta_{i},{\bf e}){\rm Tr}[ {\bf e} \psi]+f_{ 2 }(\beta_{i},{\bf e}){\rm Tr}[ {\bf e}{\bf e} \psi]+f_{ 3 }(\beta_{i},{\bf e}){\rm Tr}[{\bf e} {\bf e}{\bf e} \psi]+\mathcal{O}(\psi^{2}) \ ,
\ee
where the $f_{i}$'s only depend on the vielbein ${\bf e}^{a}$ and the constant parameters $\beta_{i}$. 

 Therefore, if we can use the four remaining gauge symmetries generated by the $P^{\nu_{1}\ldots\nu_{n}}{}_{\mu}$'s to gauge fix these four interactions to zero, the gauge fixed action will be quadratic and higher order in $\psi$, $\psi_{\mu}{}^{\nu}=0$ will solve the $\psi$ equations of motion and after integrating out $\psi$ our action will coincide with the usual dRGT action written without any St\"uckelberg fields, establishing dynamical equivalence.
 
A calculation demonstrates that the transformation generated by the $P^{\nu_{1}\ldots\nu_{n}}{}_{\mu}$'s causes $\psi_{\mu}{}^{\nu}$ to transform as
\be
 \psi_{\mu}{}^{\nu}\mapsto \psi_{\mu}'{}^{\nu}= \psi_{\mu}{}^{\nu}+\frac{1}{2}\left (\partial_{\mu}c^{\nu}+\partial^{\nu}c_{\mu}\right )+\mathcal{O}(c^{2},\psi c,\psi^{2})\ .
\ee
  We can fix the desired gauge if we can choose the four independent components of $c^{\mu}$ to simultaneously satisfy the four equations
 \begin{align}
{\rm Tr}[\partial c]&=-{\rm Tr}[\psi]\nn
 {\rm Tr}[{\bf e}\partial c]&= -{\rm Tr}[{\bf e}\psi]\nn
  {\rm Tr}[{\bf e}{\bf e}\partial c]&= -{\rm Tr}[{\bf e}{\bf e}\psi]\nn
 {\rm Tr}[{\bf e}{\bf e}{\bf e}\partial c] &= -{\rm Tr}[{\bf e}{\bf e}{\bf e}\psi], \label{GaugeFixingCondition}
 \end{align}
 where $[\partial c]_{\mu}{}^{\nu}\equiv \frac{1}{2}\left (\partial_{\mu}c^{\nu}+\partial^{\nu}c_{\mu}\right )$ and $[{\bf e}]_{\mu}{}^{\nu}=e_{\mu}{}^{a}\delta_{a}^{\nu}$.  We cannot explicitly give the solution for $c^{\mu}$, but as \eqref{GaugeFixingCondition} represents four non-linear, first order PDE's, solutions for the four components of $c^{\mu}$ are expected to exist.  Assuming existence, we can then consistently set $\psi_{\mu}{}^{\nu}=0$ everywhere and the mass terms \eqref{dRGTSectionMassTerms} reduce to the original dRGT form
 \begin{align}
\mathcal{L}_{1}&  = \epsilon_{abcd}{\bf e}^{a}\wedge{\bf e}^{b}\wedge{\bf e}^{c}\wedge{\bf 1}^{d} \nn
\mathcal{L}_{2}&   =\epsilon_{abcd}{\bf e}^{a}\wedge{\bf e}^{b}\wedge{\bf 1}^{c}\wedge{\bf 1}^{d} \nn
\mathcal{L}_{3}& =\epsilon_{abcd}{\bf e}^{a}\wedge{\bf 1}^{b}\wedge{\bf 1}^{c}\wedge{\bf 1}^{d} \ .
\end{align}

 \subsubsection{A parity-violating operator}

Sticking to simple wedge products, we find that there is one more interaction we can generate which is not of the dRGT form, 
\be
\mathcal{L}' =  \eta_{\mu a}\eta_{\nu b}\Omega^{\mu}\wedge\Omega^{a}\wedge\Omega^{\nu}\wedge\Omega^{b}~,
\label{dRGTNewTrivialMassTerm}
\ee
which when gauge fixed takes the form
\be
\mathcal{L}' = \eta_{ab}\eta_{cd} {\bf 1}^a\wedge {\bf e}^b\wedge {\bf 1}^c\wedge {\bf e}^d~.
\ee

An analysis similar to that of \cite{Hinterbichler:2012cn,Andrews:2013ora} demonstrates that this parity-violating interaction is also at most linear in the lapse and the shift, and is therefore expected to be a healthy term which does not regenerate the Boulware--Deser ghost.

Unfortunately, this potentially novel mass term is trivial on the usual branch of the theory.  Writing the interaction out in components, it is given by
\begin{align}
\mathcal{L}'&\propto \rd^{4}x\, \epsilon^{\mu a\nu b}e_{\mu a}e_{\nu b}\ ,
\end{align}
where $e_{\mu}{}^{a}$ is the vielbein, so this term depends only on the anti-symmetric parts of the vielbein.  Following \cite{Hinterbichler:2012cn,Andrews:2013ora} again, we can decompose the vielbein into a Lorentz transformation times a constrained vielbein $e_{\mu}{}^{a}=e^{\omega^{a}{}_{b}}\bar{e}_{\mu}{}^{b}$, where $\omega_{ab}\equiv\eta_{aa'}\omega^{a'}{}_{b}$ is antisymmetric and $\bar e_{\mu a}\equiv \bar e_{\mu}{}^{a'}\eta_{aa'}$ is symmetric. Since the $e^{\omega^{a}{}_{b}}$ factor is a local Lorentz transformation, it cancels in the Einstein--Hilbert term (which is built out of LLT-invariant combinations of the vielbein) and only appears in the mass terms.  Expanding the dRGT mass terms and $\mathcal{L}'$ in powers of $\omega$, their structure is such that the linear pieces vanish and they all start at $\mathcal{O}(\omega^{2})$. Hence, the $\omega$ equation of motion is solved by\footnote{The ability to set $\omega=0$ is crucial for equating the vielbein and metric formulations of dRGT~\cite{Deffayet:2012zc}.} $\omega=0$ and we can replace $e_{\mu}{}^{a}$ by the symmetric vielbein $\bar{e}_{\mu}{}^{a}$ everywhere --- a replacement which causes $\mathcal{L}'$ to vanish.  However, on a non-trivial branch of the theory on which $\omega\neq 0$,  $\mathcal{L}'$ may play a non-trivial role.

\subsubsection{Other interactions}

So far we have only constructed the dRGT terms. While it is very intriguing that they are more or less the simplest terms we can write down, they do not possess any particular symmetries not shared by other terms we could write down. Indeed, we can also form non-dRGT terms using wedge products, for example:
\be
{\cal L} = \epsilon_{acef}\eta_{bd}\Omega^{ab}\wedge \Omega^{cd}\wedge\Omega^e\wedge\Omega^f =\epsilon_{acef}\eta_{bd}\boldsymbol\omega^{ab}\wedge\boldsymbol\omega^{cd}\wedge{\bf 1}^e\wedge{\bf 1}^f~,
\ee
where after the equal sign we have gone to gauge fixed form in which this term involves two powers of the spin connection. Even further, when constructing an EFT one must include all operators compatible with the symmetries of the problem and there is no physical reason to concentrate only on terms we can write as wedge products.  Generically we can define ``covariant derivatives" of Goldstone fields in the coset construction, which proceeds as follows.  We expand the MC form as $\Omega=\rd x^{\nu}\Omega_{P\nu}{}^{\mu}P_{\mu}+\ldots$, in which the coefficient $\Omega_{P\nu}{}^{\mu}$ defines a type of vielbein.  Given a set of broken generators that transform as an irreducible representation of $H$, say $Z_{a}$, the covariant derivative of the associated Goldstone fields is given by $D_{\mu}\xi^{a}$, which is defined through
\begin{align}
\Omega_{Z}{}^{a} \equiv \rd x^{\mu}\Omega_{P\mu}{}^{\nu}D_{\nu}\xi^{a}~.
\end{align}
Symmetry preserving interactions can then be formed by contracting factors of $D_{\nu}\xi^{a}$ in $H$-invariant ways.

In the case at hand we have $\Omega_{P\nu}{}^{\mu}=\Psi_{\nu}{}^{\mu}$ and if --- for instance --- we take the $Z_{a}$'s to be the internal translation generators $P_{a}$, then we obtain the covariant derivatives 
\begin{align}
D_{\mu}\phi^{a}&= (\Psi^{-1})_{\mu}{}^{\nu}\tilde\Theta_{b}{}^{a}e_{\nu}{}^{b} \ ,
\end{align}
where $e_{\nu}{}^{b}$ is the vielbein of GR.  We see that this covariant derivative prescription simply affixes the correct factors of $\Psi_{\mu}{}^{\nu}$ and $\tilde\Theta_{a}{}^{b}$ to the vielbein $e_{\mu}{}^{a}$ such that the usual diffeomorphisms and LLT's of $e_{\mu}{}^{a}$ are translated into an $SO(1,d)_{\rm global}\times SO(1,d)_{\rm spacetime}$ rotation.

 In order to make invariant actions using the covariant derivatives we just need to contract indices using $H$-invariant tensors.  For example, staying with $D_{\mu}\phi^{a}$ we could build potentials from polynomials of $\eta^{\mu\nu}\eta_{ab}D_{\mu}\phi^{a}D_{\nu}\phi^{b}$.  Going to gauge fixed form, we find that these are nothing but potentials built from $\eta^{\mu\nu}g_{\mu\nu}$.
 
 More generally we will be able to construct arbitrary potentials built from metric fluctuations $h_{\mu\nu}=g_{\mu\nu}-\eta_{\mu\nu}$ with only the requirement that we contract indices with $\eta^{\mu\nu}$.  Therefore, the symmetry breaking pattern is not restrictive enough to single out the dRGT potentials as the unique, generic interactions which control the low energy EFT.  As discussed previously, this is the expected outcome, as it is known that loop corrections from the dRGT interactions generate terms which are not of the dRGT form, indicating that there are other interactions which share any retained symmetries of dRGT (provided symmetries are not broken by the regulators used in the quantum calculation) \cite{deRham:2013qqa}.

\section{Bi-gravity and multi-metric theories}
\label{biandmultig}

Finally, we note that our formalism extends in a straightforward way to the case of the multi-vielbein theories presented in~\cite{Hinterbichler:2012cn} (further explored in~\cite{Hassan:2012wt,Noller:2013yja,Afshar:2014dta,Scargill:2014wya}), of which ghost-free bigravity is a limiting case.  Below we provide a sketch of the construction.

\subsection{Symmetries and algebras}

 In order to construct a theory of $\mathcal{N}$ interacting vielbeins we take the group of symmetries, $G$, to be
 \be
 G=\prod_{i=1}^{\mathcal{N}} \left[ ISO_{(i)}(1,d)_{\rm local}\times {\rm Diff}_{(i)}(d+1)\right] ~,
 \ee
 which contains $\mathcal{N}$ commuting factors of our local Poincar\'e algebras generated by \\
 $\{P_{(i)}^{\mu ^{(i)}_{1}\ldots\mu ^{(i)}_{n}}{}_{a^{(i)}}, J_{(i)}^{\mu ^{(i)}_{1}\ldots\mu ^{(i)}_{n}}{}_{a^{(i)}b^{(i)}}\}$, $i\in\{1,\ldots,\mathcal{N}\}$, and $\mathcal{N}$ commuting factors of the diffeomorphism group generated by $P^{\nu^{(i)}_{1}\ldots\nu^{(i)}_{n}}{}_{\mu ^{(i)}}$, with $a^{(i)},\mu^{(i)}\in\{0,1,\ldots ,d\}$.  Any two generators indexed by $i$ and $j$ will commute when $i\neq j$.  When $i=j$ their commutation relations will simply be those given by \eqref{PoincareAlgebraLocal} and \eqref{IntercommutingAlgebra}.

\subsection{Interacting theories}
In order to construct the ghost-free, multi-vielbein interactions of \cite{Hinterbichler:2012cn} we need to choose a different preserved group $H$.  There, the studied interactions were of the form (sticking to $d+1=4$)
\begin{align}
  \mathcal{L}\sim \epsilon_{abcd}{\bf e}^{a}_{(  i )}\wedge{\bf e}^{b}_{(  j )}\wedge{\bf e}^{c}_{(  k )}\wedge{\bf e}^{d}_{(  l )}~,
  \end{align}
  for arbitrary choices of $i,j,k,l$.  Any choice of $i,j,k,l$ will lead to an action with the proper primary constraints necessary for ghost freedom.  These interaction preserve a diagonal LLT group and an independent diagonal diffeomorphism group and hence the appropriate choice of preserved subgroup is 
  \begin{align}
  H=\left (\prod_{i=1}^{\mathcal{N}}SO_{(i)}(1,d)_{\rm global}\right )_{\rm diag}\times \left (\prod_{i=1}^{\mathcal{N}}GL_{(i)}(d+1)\right )_{\rm diag}~,
  \end{align}
where the first factor is generated by $\frac{1}{2}\sum_{i=1}^{\mathcal{N}}J_{(i)a^{(i)}b^{(i)}}$ and the second by $\sum_{i=1}^{\mathcal{N}}P_{(i)}^{\mu^{(i)}}{}_{\nu^{(i)}}$.  The resulting MC form components will be the analogues of \eqref{dRGTMCComponentsStuckelberged} in highly St\"uckelberged form.

This construction now allows us to contract \textit{any} pair of upper and lower Greek indices together, irrespective of their $i$ labels, and we can contract any two Latin indices together as long as we contract using any of the $\left (\prod_{i=1}^{\mathcal{N}}SO_{(i)}(1,d)_{\rm global}\right )_{\rm diag}$ invariant tensors, i.e. $\eta$ or $\epsilon$ with any combination of indices.  For brevity, we can then drop the $i$ labels on the Latin and Greek indices in this construction, since they no longer represent important distinctions.

For our purposes here, it is easiest simply to work in gauge fixed form in which all of the fields analogous to $\tilde\Theta_{b}{}^{a}$ and $\Psi_{\mu}{}^{\nu}$ in \eqref{dRGTMCComponentsStuckelberged} are set to zero.  Having done so, the only rule for constructing actions is that they must still obey the preserved \textit{global} symmetry, i.e. $H_{\rm global}\subset H$  defined by
  \begin{align}
 H_{\rm global}&=\left (\prod_{i=1}^{\mathcal{N}}SO_{(i)}(1,d)_{\rm global}\right )_{\rm diag} \times \left (\prod_{i=1}^{\mathcal{N}}SO(1,d)_{\rm spacetime}\right )_{\rm diag}~.
  \end{align}

\subsubsection{Bi-gravity}

Specializing to the case of $\mathcal{N}=2$ in $d+1=4$ we arrive at the case of four-dimensional bi-gravity in which we give an Einstein-Hilbert term to each of the vielbeins and construct ghost-free interactions using contractions with $\epsilon$.  Explicitly, in the notation of \cite{Hinterbichler:2012cn}, the standard ghost-free bi-gravity action in $d+1=4$ is given by
\begin{align}
S&=\frac{M_{g}^{2}}{2}\int\rd^{4}x\, \det(e_{(1)})\, R[e_{(1)}]+\frac{M_{f}^{2}}{2}\int\rd^{4}x\, \det(e_{(2)})\, R[e_{(2)}] -\frac{m^{2}M_{fg}^{2}}{8}\int\mathcal{L}_{\rm int} \ ,\label{BigravityAction}
\end{align}
where $M_{fg}^{2}=1/(M_{g}^{-2}+M_{f}^{-2})$ and
\begin{align}
\mathcal{L}_{\rm int}&=\frac{\beta_{0}}{4!}\epsilon_{abcd}{\bf e}_{(1)}^{a}\wedge{\bf e}_{(1)}^{b}\wedge{\bf e}_{(1)}^{c}\wedge{\bf e}_{(1)}^{d}+\frac{\beta_{1}}{3!}\epsilon_{abcd}{\bf e}_{(2)}^{a}\wedge{\bf e}_{(1)}^{b}\wedge{\bf e}_{(1)}^{c}\wedge{\bf e}_{(1)}^{d}\nn
&\quad +\frac{\beta_{2}}{4}\epsilon_{abcd}{\bf e}_{(2)}^{a}\wedge{\bf e}_{(2)}^{b}\wedge{\bf e}_{(1)}^{c}\wedge{\bf e}_{(1)}^{d}+\frac{\beta_{3}}{3!}\epsilon_{abcd}{\bf e}_{(2)}^{a}\wedge{\bf e}_{(2)}^{b}\wedge{\bf e}_{(2)}^{c}\wedge{\bf e}_{(1)}^{d}\nn
&\quad +\frac{\tilde\beta_{0}}{4!}\epsilon_{abcd}{\bf e}_{(2)}^{a}\wedge{\bf e}_{(2)}^{b}\wedge{\bf e}_{(2)}^{c}\wedge{\bf e}_{(2)}^{d}\ .\label{BigravityInteractions}
\end{align}
The two vielbeins define the two metrics $g={\bf e}_{( 1 )}^{a}{\bf e}_{( 1 )}^{b}\eta_{ab}$ and $f={\bf e}_{( 2 )}^{a}{\bf e}_{( 2 )}^{b}\eta_{ab}$, each with its own corresponding Planck mass, $M_{g}$ and $M_{f}$, respectively.

Working in gauge fixed form, we see that indeed all of these forms can be created in our present construction.  Moreover, considering wedge products, there is an additional parity violating term we can generate which is consistent with all the required symmetries.  In gauge fixed form, this is
\begin{align}
\mathcal{L}'\sim \eta_{ab}\eta_{cd}{\bf e}_{(  1 ) }^{a}\wedge{\bf e}_{(  2 ) }^{b}\wedge{\bf e}_{(  1 ) }^{c}\wedge{\bf e}_{(  2 ) }^{d} \ ,
\end{align}
which is nothing but the analogue of the trivial mass term found in the dRGT case \eqref{dRGTNewTrivialMassTerm}.  Again, the analysis of \cite{Hinterbichler:2012cn,Andrews:2013ora} demonstrates that this parity-violating term $\mathcal{L}'$ is at most linear in the lapses and shifts of the two vielbeins and hence potentially ghost-free.

In the dRGT case we found that the $\mathcal{L}'$ term was trivial on the normal branch of the theory, and we will find the same here.  Given our two vielbeins $e_{(1)\mu}{}^{a}$ and $e_{(2)\mu}{}^{a}$ we consider their contraction in the form $e_{(1)}^{\mu}{}_{a}e_{(2)\mu b}$ and then decompose the first vielbein, say, as $e_{(1)}^{\mu}{}_{a}=e^{\omega_{a}{}^{c}}\bar e_{(1)}^{\mu}{}_{c}$, now chosen such that the combination $\bar e_{(1)}^{\mu}{}_{c}e_{(2)\mu b}$ is symmetric under $b\leftrightarrow c$.  
We can rewrite our term as
\begin{align}
\mathcal{L}'&\sim \det(e_{(1)})\epsilon^{abcd}e_{(1)}^{\nu}{}_{a}e_{(2)\nu b}e_{(1)}^{\mu}{}_{c}e_{(2)\mu d}~,
\end{align}
and so if we expand our decomposition in powers of $\omega$ the expansion starts at $\mathcal{O}(\omega^{2})$. Therefore $\omega=0$ remains a solution, and on this branch $\mathcal{L}'$ vanishes.  Again, if there exist branches of the theory on which $\omega\neq 0$ then $\mathcal{L}'$ may play an important role.

\subsubsection{Multi-vielbein Theories}

The case of $\mathcal{N}>2$ interacting vielbeins in $d+1=4$ proceeds similarly.  Again, each vielbein will acquire its own Einstein-Hilbert and cosmological constant terms and the fields can interact through potentials constructed by wedging together vielbeins\footnote{There are, of course, other interactions which obey the relevant symmetries, but we choose to study only these special potentials which lead to our desired primary constraints.} and contracting Latin indices with either $\epsilon$ or $\eta$.

 First, consider generating potentials by only using the bi-gravity interactions of the form \eqref{BigravityInteractions}, so that there are at most two distinct vielbeins interacting at a given vertex. Following \cite{Hinterbichler:2012cn}, we can depict these theories by drawing a node for each type of vielbein and drawing a line between any two nodes whose corresponding vielbeins interact via a bi-gravity vertex.
 
 Any theory constructed in this manner whose depiction is free of closed loops is known to be equivalent to a metric theory on any branch in which the bi-gravity type symmetry condition $e^{\mu}_{(i) [a}e_{(j)|\mu |b]}=0$ holds, for any two vielbeins which interact with each other.  This is the case independent of the parameters appearing in front of the bi-gravity type potentials.  For such theories, we can consider adding our parity violating operators of the form $\sim \eta_{ab}\eta_{cd}{\bf e}_{( i) }^{a}\wedge{\bf e}_{( j) }^{b}\wedge{\bf e}_{( i) }^{c}\wedge{\bf e}_{( j) }^{d}$, but again they vanish on the normal branch where we impose the symmetric vielbein conditions $e^{\mu}_{(i) [a}e_{(j)|\mu |b]}=0$.
  
 If there are closed loops in the depiction of the theory, or if more than two types of vielbeins interact at a given vertex, however, it is no longer consistent to impose the symmetric vielbein conditions and instead the restrictions we impose depend on the details and parameters of the interactions used.  In this case, our parity violating operators may no longer vanish on the primary branches of the theory and may be as important as any of the other constraint preserving interactions.

 \section{Conclusions}
 \label{conclusion}

The dRGT theory of massive gravity provides the optimal ({\it i.e.}, highest strong coupling scale) low energy starting point for any putative theory of General Relativity in a Lorentz invariant Higgs phase on flat space.  Ideally, one should seek an explicit microphysical model that connects the high and low energy endpoints of the gravitational sector through spontaneous symmetry breaking via associated Higgs fields in a manner analogous to the breaking of Yang--Mills theories.  However, such a construction has remained elusive and we therefore turn to more oblique studies.

In this paper, we have approached massive gravity (and in particular dRGT) through the general framework of nonlinear realizations.  Such a study requires an understanding of how to apply coset techniques to gauge theories and in exploring spontaneously broken GR we have relied heavily on previous work \cite{Goon:2014ika} which developed the implementation of these methods with respect to spontaneously broken Yang--Mills theory.

The study of non-linear realizations has a long history, as can be seen in the extensive, yet incomplete, set of references\cite{Pashnev:1997xk,Zumino:1970tu,Riccioni:2009hi,Borisov:1974bn, Isham:1971dv,Ogievetsky:1974,McArthur:2010zm,Low:2001bw,deAzcarraga:1998uy,Coleman:1969sm, D'Hoker:1994ti,Weinberg:1968de, Ivanov:1975zq,Ogievetsky:1973ik,Ivanov:1976pg,Brauner:2014aha,volkov,Goon:2012dy,Delacretaz:2014oxa,Goon:2014ika, Kirsch:2005st,Boulanger:2006tg,Callan:1969sn,Ivanov:1981wn,Volkov:1973vd}, and correspondingly some facets of our work echo previous results in the literature.  We do not claim that the \textit{general} methods employed here are entirely novel; indeed the use of coset techniques in studying gauge theories where spacetime symmetries play an important role can be found in several of the quoted references above.  Instead, it is our specific constructions and applications that are new.  Our particular formulation of General Relativity in coset language has the conceptual advantage that \textit{both} local Lorentz transformations and diffeormorphisms have equal footing as non-linear symmetries of the system, whereas (to our knowledge) previous constructions only focus on one of the two groups.  This conceptual advantage then turns into a \textit{technical} one when broken phases of gravity are considered.  It is essential for constructing dRGT-like massive gravity theories that the residual symmetry group becomes a diagonal subgroup of LLT's and diffeomorphisms, as is made explicit through our methods.  The steps which lead from the construction of GR to that of a dRGT phase would be much less straightforward in any construction different from ours.

We have focused on the search for the most highly restrictive symmetry breaking pattern ({\it i.e.}, the pattern with the most retained symmetries) whose low energy theory will admit terms of the dRGT form.  Once found, this provides a systematic method for determining the allowed interactions of the effective field theory and can help guide future searches for microphysical models.

The appropriate symmetry breaking pattern we found is to preserve the diagonal subgroup consisting of global Lorentz transformations and the corresponding diffeomorphisms which also generate a global $SO(1,d)$ transformation.  Further, this was demonstrated to be the largest possible preserved subgroup consistent with our requirement that the dRGT terms appear in the low energy action.  As expected, there are additionally a host of other non-derivative interactions (and higher derivative interactions, see \cite{Folkerts:2011ev,Hinterbichler:2013eza,deRham:2013tfa,Kimura:2013ika,deRham:2015rxa,Noller:2014ioa} for more on the possibility of ghost-free higher derivative terms in massive gravity) which obey the required symmetry and are not of the dRGT form.

Among these non-dRGT potentials is the parity violating operator
\begin{align}
\mathcal{L}'&\sim \eta_{ab}\eta_{cd}{\bf e}^{a}\wedge{\bf 1}^{b}\wedge{\bf e}^{c}\wedge{\bf 1}^{d} \ ,
\end{align}
the wedge structure of which ensures that it is at most linear in the lapse and shift, and hence preserves the constraint structure required to remove the Boulware--Deser ghost.  However, the normal branch of the theory enforces the symmetry condition $e_{[\mu a]}=0$ which causes this interaction to vanish. This parity violating term could still play an important role on non-trivial branches of the theory.

Finally, we have extended our methods to the cases of bi-gravity and multi-vielbein models.  These are natural extensions of the coset construction of dRGT and we are able to reproduce the theories of \cite{Hinterbichler:2012cn}.  New parity violating interactions were again found in both of these cases, but we leave it to future work to determine whether these have non-trivial effects for any classes of the multi-vielbein theories.

\noindent
{\bf Acknowledgments:} We thank Riccardo Penco and Rachel Rosen for useful discussions.
GG gratefully acknowledges support from a Starting Grant of the European Research Council (ERC StG grant 279617).  Research at Perimeter Institute is supported by the Government of Canada through Industry Canada and by the Province of Ontario through the Ministry of Economic Development and Innovation.  This work was made possible in part through the support of a grant from the John Templeton Foundation. The opinions expressed in this publication are those of the author and do not necessarily reflect the views of the John Templeton Foundation (KH).
This work was supported in part by the Kavli Institute for Cosmological Physics at the University of Chicago through grant NSF PHY-1125897, an endowment from the Kavli Foundation and its founder Fred Kavli, and by the Robert R. McCormick Postdoctoral Fellowship (AJ). The work of MT was supported by the US Department of Energy grant DE-FG02-95ER40893.

\appendix

\section{\label{Appendix:GLConnection}The $GL(d+1)$ Connection and Related Terms}

The preserved subgroup in the construction of GR was $SO(1,d)_{\rm global}\times GL(d+1)$.  Each factor in the preserved subgroup has an associated connection, given by the component of the Maurer-Cartan form along the relevant generator.  For the $SO(1,d)_{\rm global}$ factor, this is the familiar spin connection, after appropriate inverse Higgs constraints have been enforced.  The connection for the $GL(d+1)$ factor was ignored, however, as it does not end up playing a role in the construction of GR.  In this appendix, we explore this ignored connection and related quantities for completeness.

The $GL(d+1)$ connection is the component of the Maurer-Cartan form which lies along the $P^{\mu}{}_{\nu}$ generator.  The coset element for GR \eqref{CosetElementDiffsAndLLTs} contains an infinite number of factors, but only a few are relevant for the present calculation.  It is sufficient to work with 
\begin{align}
g=e^{x^{\mu}P_{\mu}}e^{\chi_{\mu\nu}{}^{\rho}P^{\mu\nu}{}_{\rho}}\left (\ldots\right )
\end{align}
which leads to
\begin{align}
g^{-1}\rd g&=\rd x^{\mu}P_{\mu}-2\chi_{\alpha\beta}{}^{\nu}\rd x^{\alpha}P^{\beta}{}_{\nu}
\end{align}
and hence the $GL(d+1)$ connection one-form is simply given by $\Omega _{\beta}{}^{\nu}\equiv-2\chi_{\alpha\beta}{}^{\nu}\rd x^{\alpha}$.  This cannot be rewritten in terms of other fields via inverse Higgs constraints.  The relevant commutator to examine is $[P_{\mu},P^{\lambda}{}_{\nu}]=-\delta_{\mu}^{\lambda}P_{\nu}$, but setting the right side to zero is nonsensical as it sends $x^{\mu}\to 0$ and in any case such a replacement would not lead to an identification between $\chi_{\alpha\beta}{}^{\nu}$ and any other fields in the theory.

The connection itself only appears in the covariant derivative used to couple external matter fields which transform under $GL(d+1)$ to the fields used for the non-linear realization.  However, $\chi_{\alpha\beta}{}^{\nu}$ has no dynamics on until we give it a kinetic term which requires finding higher order pieces in the Maurer-Cartan expansion, i.e. those along $P^{\mu\nu}{}_{\sigma}$ and similar generators with more Greek indices.  These correspond to the curvature tensor associated to the connection $\omega'$ and covariant derivatives thereof, after inverse Higgs.

All such terms have been ignored in the text because they they are difficult to build actions from and they do not end up affecting the physics anyway. Apart from matter covariant derivatives which involve $\Omega_{\beta}{}^{\nu}$, the only building blocks at our disposal which have only Greek indices are $\Omega^{\mu}$, $\Omega_{\alpha\beta}{}^{\mu}$, $\Omega_{\alpha\beta\gamma}{}^{\mu}$, etc.  Attempting to build invariant actions using \textit{only} these pieces, we quickly find ourselves to be very limited.    The residual $GL(d+1)$ symmetry forces us to contract each upper Greek index with a lower Greek index (raising and lowering of Greek indices is not allowed) and we find that there are only two possible ways to wedge together any of the above terms to create a 4-form whose indices are all appropriately contracted,
\begin{align}
\mathcal{L}\sim\begin{cases} \Omega^{\mu}\wedge\Omega_{\mu\nu}{}^{\nu}\wedge\Omega^{\alpha}\wedge\Omega_{\alpha\beta}{}^{\beta}\\ {\rm or} \\ \Omega^{\mu}\wedge\Omega_{\mu\nu}{}^{\beta}\wedge\Omega^{\alpha}\wedge\Omega_{\alpha\beta}{}^{\nu}\end{cases}\ .
\end{align}
Both are $\mathcal{O}({\rm curvature}^2)$ and considered to be subleading corrections, though it is possible that further study of these terms could be interesting.   If we then consider mixed terms which also use components with Latin indices, then we can try wedging the above ingredients together with $\Omega^{a},\Omega_{\mu}{}^{a},\Omega_{\mu}{}^{ab}$, etc. and we'll find that the only invariants we can make are those which already appeared in the GR construction, such as $\rd^{4}x\, \sqrt{-g}R\propto \epsilon_{abcd}\Omega^{a}\wedge\Omega^{b}\wedge\Omega^{\mu}\wedge\Omega_{\mu}{}^{cd}$. No such invariants are constructable using $\Omega_{\alpha\beta}{}^{\nu}$ or other terms with more indices.  Since these are the only terms involving dynamical fields ($\Omega^{\mu}$ only contains the coordinate $x^{\mu}$), we see that $\chi_{\alpha\beta}{}^{\nu}$ and related fields never couple to the vielbein or spin connection.  As we're only interested in the latter quantities, we ignore the $GL(d+1)$ connection and associated quantities in the text.

\renewcommand{\em}{}
\bibliographystyle{utphys}
\bibliography{dRGTCosetJHEPEdited3}

\end{document}